\documentclass[pra,twocolumn,showpacs,preprintnumbers,amsmath,amssymb]{revtex4}

\usepackage{graphicx}
\usepackage{dcolumn}
\usepackage{bm}

\def \ed {\end{document}}
\def\Fbox#1{\vskip1ex\hbox to 8.5cm{\hfil\fboxsep0.3cm\fbox{%
  \parbox{8.0cm}{#1}}\hfil}\vskip1ex\noindent}  

\def\be{\begin{equation}}\def\ee{\end{equation}}
\def\bea{\begin{eqnarray}}\def\eea{\end{eqnarray}}
\def\bse{\begin{subequations}}\def\ese{\end{subequations}}
\newcommand{\BE}[1]{\begin{equation}\label{#1}}
\newcommand{\BEA}[1]{\begin{eqnarray}\label{#1}}
\newcommand{\BSE}[1]{\begin{subequations}\label{#1}}

\let \= \equiv \let\*\cdot \let\~\widetilde \let\^\widehat \let\-\overline

\def\<{\left\langle}    \def\>{\right\rangle}
\def\({\left(}          \def\){\right)}
 \def \[ {\left [} \def \] {\right ]}

\begin{document}

\title{Counterflow instability and turbulence in a spin-1 spinor Bose--Einstein condensate}

\author{Kazuya Fujimoto $^1$ and Makoto Tsubota $^{1,2}$}
\affiliation{$^1$Department of Physics, Osaka City University, 3-3-138 Sugimoto, Sumiyoshi-ku, Osaka 558-8585, Japan   \\
$^2$ The OCU Advanced Research Institute for Natural Science and Technology (OCARINA),Osaka City University, 3-3-138 Sugimoto, Sumiyoshi-ku, Osaka 558-8585, Japan}

\date{\today}

\begin{abstract}
We theoretically study counterflow instability and turbulence in a spin-1 spinor Bose--Einstein condensate by the Gross--Pitaevskii equation and the Bogoliubov--de Gennes equation. Our study considers (i) the dynamics induced by the counterflow of two components with different magnetic quantum numbers, which 
leads to turbulence with spin degrees of freedom, and (ii) the properties of the turbulence. For (i), the behavior of the condensate induced by the counterflow strongly depends on whether the spin-dependent interaction is ferromagnetic or antiferromagnetic, leading to 
different behaviors for the dispersion relation and the spin density vector, $etc$. 
For (ii), we numerically calculate the spectrum of the spin-dependent interaction energy, which also depends on the spin-dependent interaction. The spectrum of the spin-dependent interaction energy in the ferromagnetic case exhibits a $-7/3$ power law, whereas that in the antiferromagnetic case does not. 
The $-7/3$ power law can be explained by scaling analysis.
\end{abstract}


\maketitle

\section{INTRODUCTION}

Turbulence is one of the most important topics in modern physics. Turbulence in classical fluids has been studied for a long time \cite{Frisch}, but turbulence is also observed in diverse fields such as low-temperature physics, plasma physics, and astronomy, $etc$. 
Kolmogorov studied turbulence in classical fluids in 1941 \cite{K41} and significantly advanced the study of turbulence. 
He found that the kinetic energy spectrum obeyed a $-5/3$ power law, which became known as the Kolmogorov $-5/3$ power law. 

Several methods have been used for generating turbulence in classical fluids. Reynolds performed a famous experiment in 1883 in which he
created turbulence by using flow through a circular pipe. Another method uses hydrodynamic instabilities, which have been investigated for a long 
time. There are many kinds of instabilities such as Kelvin--Helmholtz (KH) and Rayleigh--Taylor (RT) instabilities \cite{Chandrasekhar} which can lead to turbulence. 

Hydrodynamic instability has recently been studied in atomic Bose--Einstein condensates (BECs). Atomic BECs exhibit different dynamics from classical fluids 
because they are quantum fluids. Quantized vortices occur in BECs; these vortices are nucleated through the hydrodynamic instability.  KH and RT instabilities have been studied in two-component atomic BECs. For KH instability, quantized vortices nucleate at the boundary layer between the two components \cite{Takeuchi10}, whereas RT instability 
generates the shape of mushroom in the condensate winded by the quantized vortex ring  \cite{Sasaki09}. Moreover, counterflow instability in two-component atomic BECs results in vortices that nucleate by the collapse of solitons \cite{Ishino10, Ishino11}. 
Thus, hydrodynamic instability has been extensively investigated in two-component atomic BECs. 

Unlike two-component atomic BECs, spinor BECs have spin degrees of freedom and exhibit phenomena characteristic of spin. 
Collisions of spin-1 spinor BECs have been investigated numerically and show different results from one- and two-component BECs \cite{Guilleumas08}. 
The hydrodynamic equation for spinor BECs has recently been studied \cite{Lamacraft08, Kudo10, Kudo11}. 
However, hydrodynamic instability in spinor BECs has been investigated less than in two-component BECs.
We expect that hydrodynamic instability in spinor BECs will exhibit unique behavior due to their spin degrees of freedom and form turbulent states in which the spin density vector has different directions.

Quantum turbulence has been investigated for a long time in superfluid $^{4}\rm{He}$ and $^{3}\rm{He}$ \cite{PLTP}, being currently studied in atomic BECs too \cite{Henn09, Seman11, Kobayashi07, Horng09}. 
Numerical studies predict that the Kolmogorov $-5/3$ power law, which was first observed 
in classical fluids, also holds in atomic BECs \cite{Kobayashi07}. Turbulence in two-component BECs has been investigated \cite{Ishino10,Ishino11}. We expect that the turbulent 
state in a spin-1 spinor BEC will exhibit properties characteristic of the spin degrees of freedom, which one- and two-component BECs do not show. This is one of the major themes of this paper. 

In this paper, we focus on counterflow instability in spin-1 spinor BECs in a homogeneous two-dimensional system. 
There are three reasons for studying the instability in this system. 
The first reason is that the dynamics of spin-1 spinor BECs induced by counterflow exhibit characteristic behaviors. 
Dynamics peculiar to counterflow have been observed in two-component BECs \cite{Ishino10, Ishino11}. However, there are distinct differences between 
two-component BECs and spinor BECs. 
A spinor BEC has spin degrees of freedom.
The number of particles of each component is conserved in a two-component BEC, whereas it is not conserved in a spinor BEC because of the spin-dependent interaction.
Therefore, we expect that a spin-1 spinor BEC will exhibit dynamics characteristic of not only the counterflow but also the spin degrees of freedom. 
The second reason for studying counterflow instability in spin-1 spinor BECs is that counterflow instability can generate turbulence in spin-1 spinor BECs.
As stated above, one aim of this study is to investigate the behavior of turbulence in spin-1 spinor BECs. 
The third reason is that counterflow in two-components BECs has been experimentally investigated \cite{Hamner11,Hoefer11}. 
We expect that it may be possible to experimentally study counterflow of spin-1 spinor BECs.
Hence, we study the counterflow instability of spin-1 spinor BECs using the Gross--Pitaevskii (GP) equation and the Bogoliubov--de Gennes (BdG) equation.
We consider counterflow between the $m = 1$ and $m = -1$ components, where $m$ is the magnetic quantum number. 

Our main purposes are to investigate phenomena characteristic of the spin degrees of freedom induced by the counterflow instability and the properties of a turbulent state in spin-1 spinor BECs. 
For the instability, we investigate the pattern of the particle number density and the time dependence of the magnitude of the spin density vector, $etc$. 
For the turbulence of a spin-1 spinor BEC, we focus on statistical quantities such as the probability density function (PDF) of the magnitude of the spin density vector and 
the spectrum of the spin-dependent interaction energy. 

The results obtained reveal that the behaviors of the instability and the turbulent state induced by the counterflow depend greatly on the sign of the spin-dependent interaction. 
We calculate the dispersion relation obtained from the BdG equation, the PDF of the magnitude of 
the spin density vector and the spectrum of the spin-dependent interaction energy; these quantities exhibit different behaviors depending on 
whether the spin-dependent interaction is ferromagnetic or antiferromagnetic.
As for the spectrum of the spin-dependent interaction energy, the $-7/3$ power law is clearly found in the ferromagnetic case, but not in the antiferromagnetic case. 
The $-7/3$ power law can be understood by the scaling analysis of the time development equation of the spin density vector. 

This paper is organized as follows. Section II describes the formulation. In Sec. III, we analytically calculate the BdG equation. Section IV presents the numerical 
results related to the dynamical instability induced by counterflow. The turbulent state in a spin-1 spinor BEC is treated in Sec. V. 
In Sec. VI, we discuss some problems of our study. Finally, we summarize the findings in Sec. VII.

\section{FORMULATION}

\subsection{Gross--Pitaevskii equation}

We consider a spin-1 spinor BEC in a homogeneous two-dimensional system at zero temperature because this system is easy to study theoretically and
is well described by macroscopic wave functions $\psi _m$ ($m = 1,0,-1$). 
For simplicity, we do not treat a magnetic field or the dipole--dipole interaction. The wave functions $\psi _m$ then obey 
the GP equation \cite{Ohmi98, Ho98}:

\begin{equation}
i\hbar \frac{\partial}{\partial t} \psi _{m} =  -\frac{\hbar ^2 }{2M} \nabla ^2 \psi _{m} + c_{0} n \psi _{m} + c_{1} \sum _{n=-1} ^{1} \bm{s} \cdot \bm{S} _{mn} \psi _{n}, 
\end{equation}
where $M$ is the mass of a particle. The total density $n$ and the spin density vector $\bm{s}$ are respectively given by

\begin{equation}
n =  \sum _{m=-1} ^{1}|\psi _m|^2, 
\end{equation}

\begin{equation}
s_{i} = \sum _{m,n = -1}^{1} \psi _{m}^{*} (S_{i})_{mn} \psi _{n}, 
\end{equation}
where $(S_{i})_{mn}$ are the spin-1 matrices: 
\begin{equation}
S_{x} = \frac{1}{\sqrt{2}}
\begin{pmatrix} 
0 & 1 & 0 \\
1 & 0 & 1 \\
0 & 1 & 0
\end{pmatrix}
,
\end{equation}
\begin{equation}
S_{y} = \frac{i}{\sqrt{2}}
\begin{pmatrix} 
0 & -1 & 0 \\
1 & 0 & -1 \\
0 & 1 & 0
\end{pmatrix}
,
\end{equation}
\begin{equation}
S_{z} = 
\begin{pmatrix} 
1 & 0 & 0 \\
0 & 0 & 0 \\
0 & 0 & -1
\end{pmatrix}
.
\end{equation}
The parameters $c_{0}$ and $c_{1}$ are the coefficients of the spin-independent and dependent interactions for two-dimensional system. 

The total energy $E$ is given by 
\begin{eqnarray}
\nonumber E &=& \int \sum _{m=-1} ^{1} [\psi ^{*} _{m} (-\frac{\hbar ^{2}}{2M}\nabla ^2) \psi _{m}] d \bm{r} \\ 
&+& \frac{c_{0}}{2} \int n^{2} d \bm{r} + \frac{c_{1}}{2} \int \bm{s} ^{2} d \bm{r}.
\end{eqnarray}
The spin-dependent interaction energy is the last term with the coefficient $c_{1}$ on the right-hand side of Eq. (7). 
The ground state in a homogeneous system without a magnetic field is ferromagnetic for $c_{1}<0$ and polar for $c_{1}>0$. 

The total particle number and the spin in the $z$ direction are conserved in the GP model without a magnetic field and the dipole-dipole interaction, namely 
\begin{equation}
\frac{d}{dt} (N_{1}+N_{0}+N_{-1}) = 0, 
\end{equation}
\begin{equation}
\frac{d}{dt} (N_{1}-N_{-1}) = 0, 
\end{equation}
with $N_{m} = \int |\psi _m|^2 d\bm{r} $. Equations (8) and (9) show that $N_{1}$ has the same time evolution as $N_{-1}$ and that the change of $N_{0}$ is related to
that of $N_{1}$ and $N_{-1}$; this is important for understanding the instability induced by the counterflow (see Sec. IV). 

\subsection{Initial state}
We consider the counterflow between the $m = 1$ and $m = -1$ components with a relative velocity $\bm{V}_{R} = V_{R} \hat{\bm{e}}_{x}$ in a homogeneous two-dimensional system, where 
$\hat{\bm{e}}_{x}$ is a unit vector along the x direction. The initial state is expressed by 
\begin{equation}
\begin{pmatrix} 
\psi _{1}^{(0)} \\
\psi _{0}^{(0)} \\
\psi _{-1}^{(0)}
\end{pmatrix}
= \sqrt{\frac{n_{0}}{2}}
\begin{pmatrix} 
{\rm{exp}}[i(\frac{M}{2 \hbar}\bm{V}_{R}\cdot \bm{r} - \frac{\mu _{1}}{\hbar}t)] \\
0 \\
{\rm{exp}}[-i(\frac{M}{2 \hbar}\bm{V}_{R}\cdot \bm{r} + \frac{\mu _{-1}}{\hbar}t)]
\end{pmatrix}
,
\end{equation}
where $n_{0}$ is the total density and $\mu _{1}$ and $\mu _{-1}$ are the chemical potentials and are equal to $c_{0}n_{0} + MV_{R}^2/8$. 
We use this initial state to investigate the counterflow instability and the turbulent state in a spin-1 spinor BEC. 

\subsection{Numerical method}
We use the Crank--Nicholson method to numerically calculate the GP equation starting from the initial state of Eq. (10). 
The coordinate is normalized by the coherence length $\xi = \hbar /\sqrt{2Mc_{0}n_{0}}$ and the box size is $128\times 128$. 
Space in the $x$ and $y$ directions is discretized into $512 \times 512$ bins. 
The time is normalized by $\tau = \hbar /c_{0}n_{0}$.
We add some small white noise to the initial state of Eq. (10); without this noise, the instability cannot be generated.
In our calculations, the noise forms the particles of the $m=0$ component, which accounts for  $0.1 \sim 0.3 \%$ of the total particle number. 
This is consistent with experimental results \cite{Sadler06}. 

\section{BdG equation and dispersion relation}
In this section, we consider a small deviation $\delta \psi _{m}$ from the initial state of Eq. (10), whose dispersion relation can be obtained by linear analysis.

\begin{figure}[t]
\begin{center}
\includegraphics[keepaspectratio, width=7cm,clip]{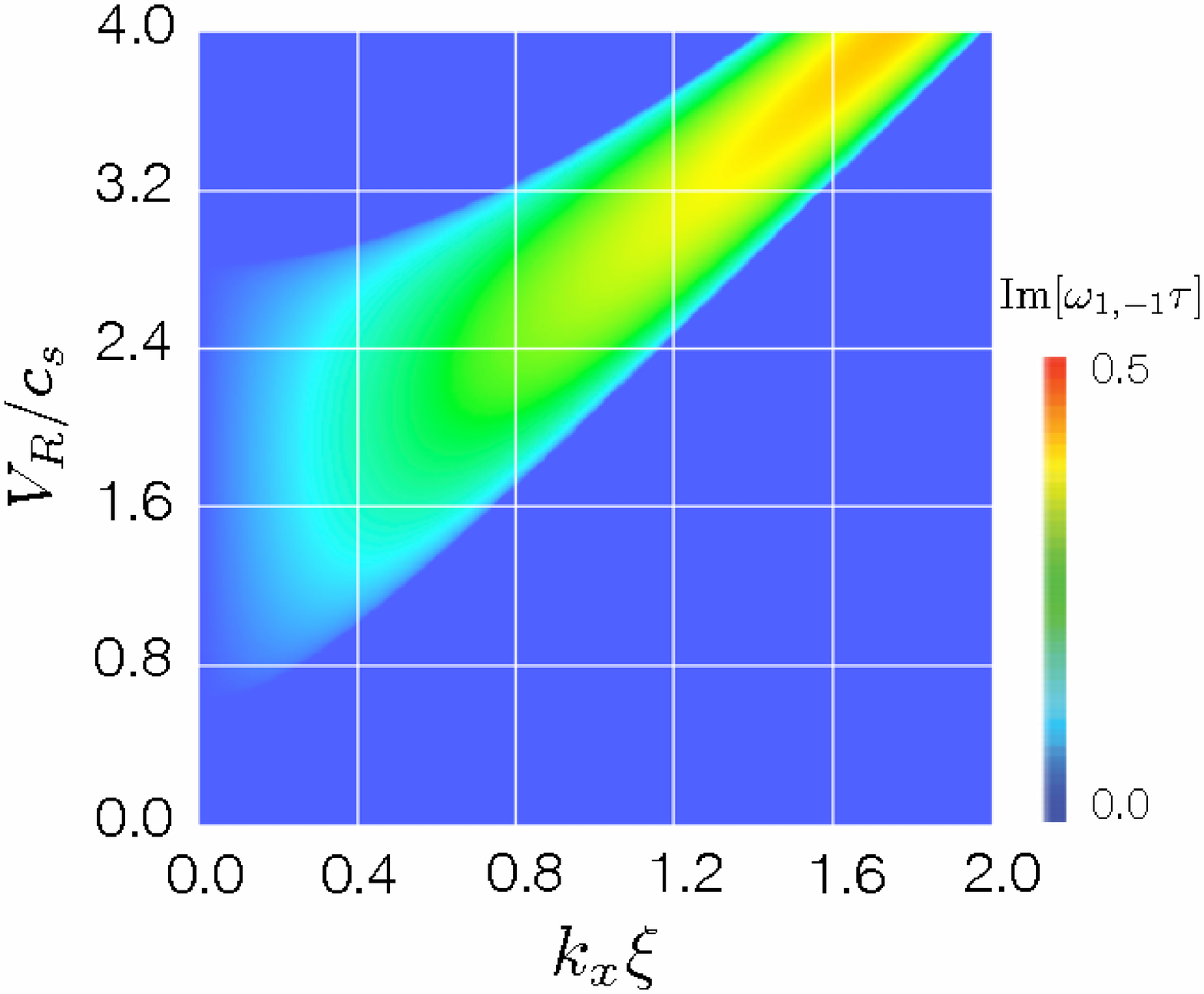}
\includegraphics[keepaspectratio, width=7cm,clip]{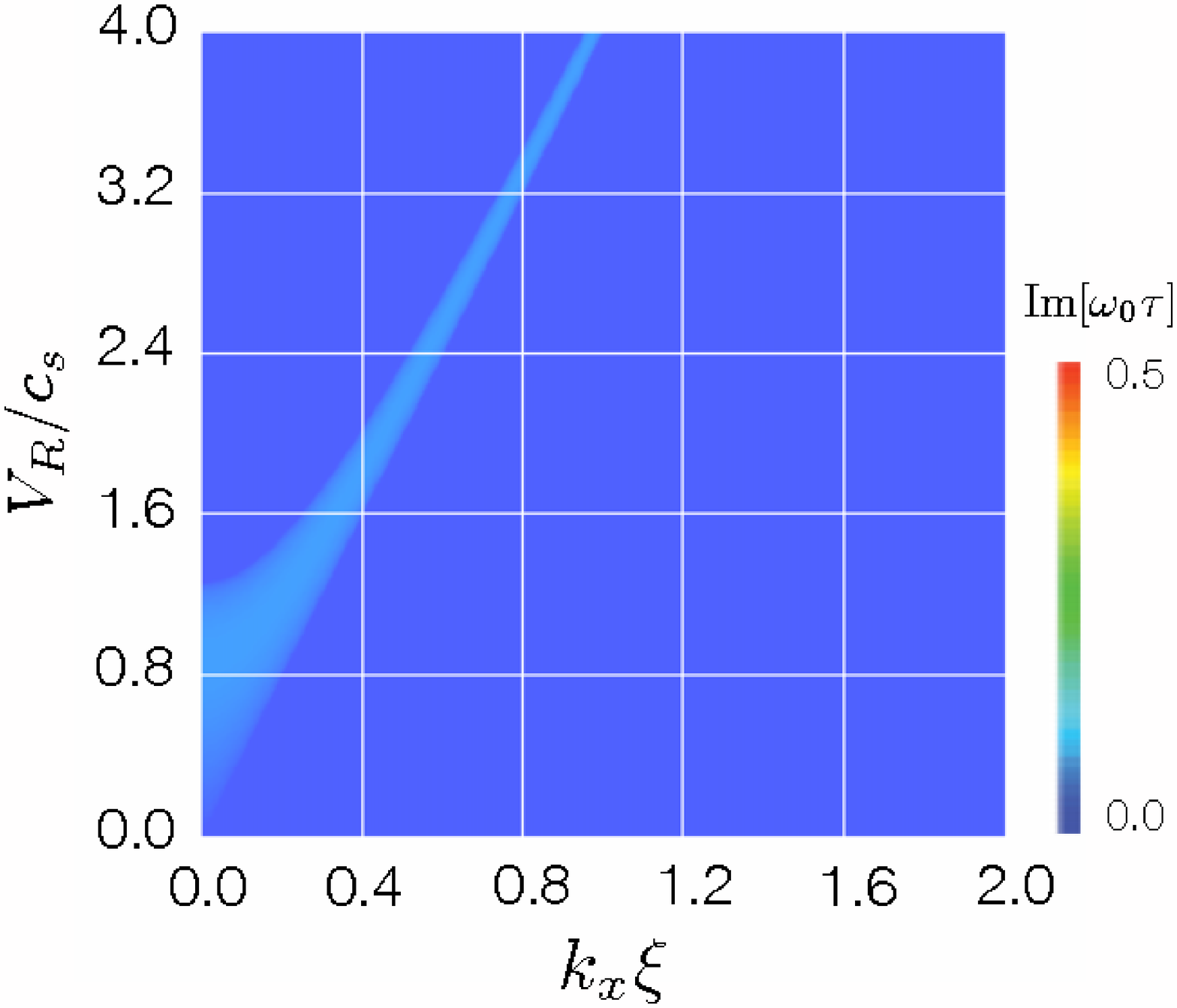}
\caption{Imaginary part of the dispersion relation for the antiferromagnetic case with $c_{0}/c_{1} = 20$ and $c_{0}>0$. 
The upper and lower graphs are respectively the imaginary parts of $\omega _{1,-1}$ and $\omega _{0}$.
The vertical and horizontal axes are respectively the relative velocity and wave number in the $x$ direction, which is normalized by the 
sound velocity $c_{s} = \sqrt{c_{0}n_{0}/2M}$ and the coherence length $\xi = \hbar /\sqrt{2Mc_{0}n_{0}}$.
The imaginary part of $\omega _{0}$ has finite values for any finite relative velocity $V_{R}$, while that of $\omega _{1,-1}$ has finite values only for a 
relative velocity larger than the critical value $V_{c}$. } 
\end{center}
\end{figure}

\begin{figure}[t]
\begin{center}
\includegraphics[keepaspectratio, width=7cm,clip]{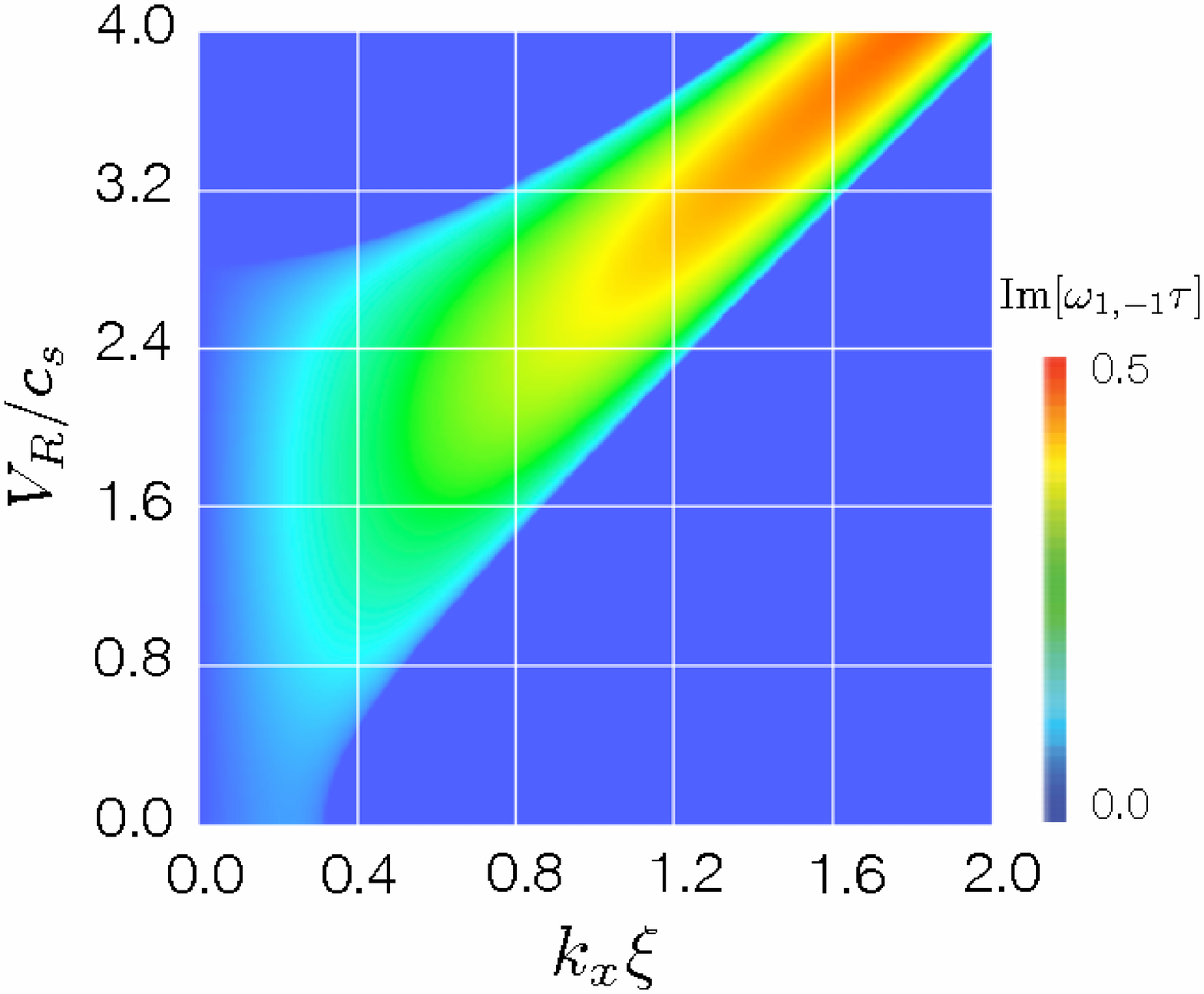}
\includegraphics[keepaspectratio, width=7cm,clip]{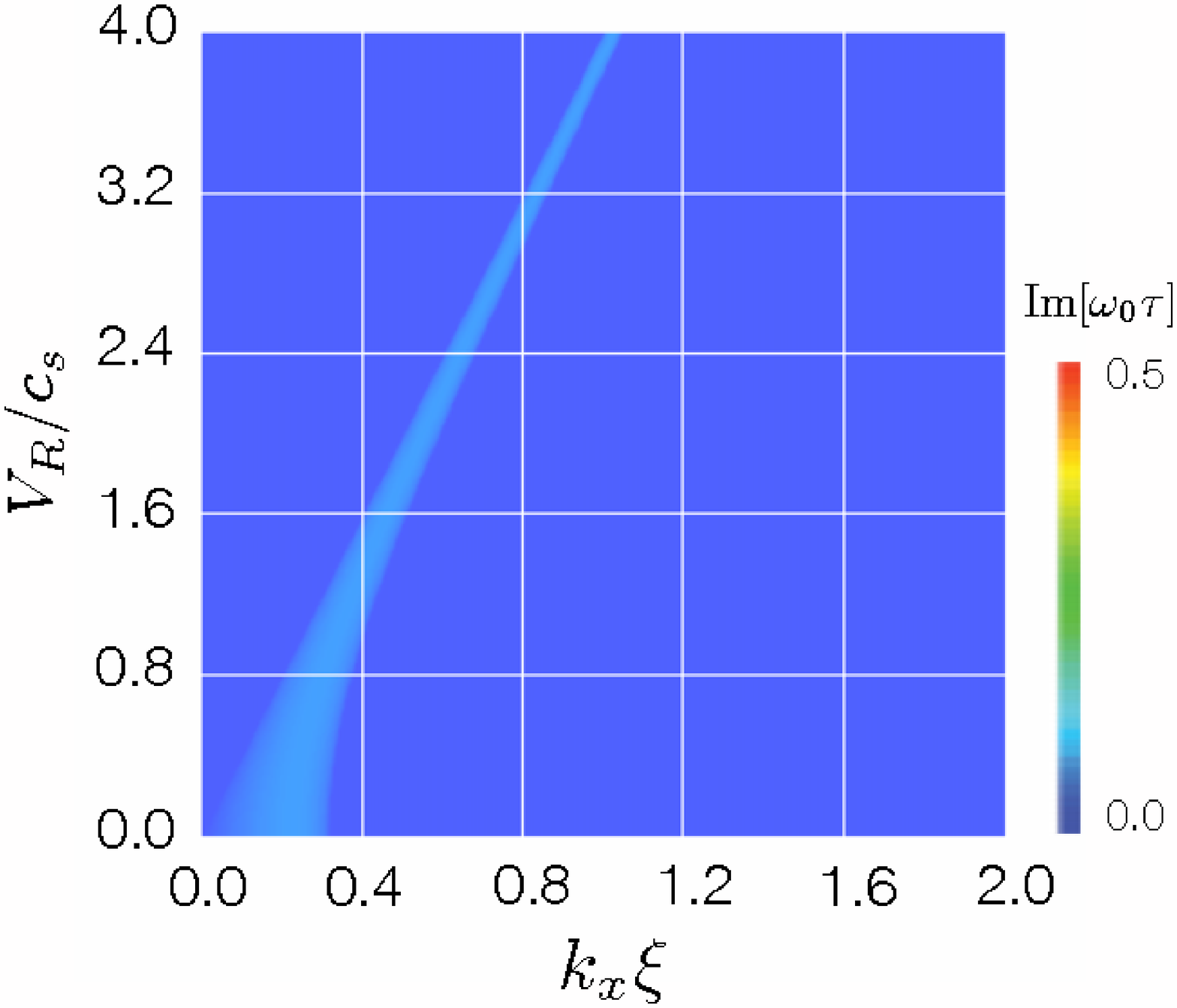}
\caption{Imaginary part of the dispersion relation for the ferromagnetic case with $c_{0}/c_{1} = -20$ and $c_{0}>0$. 
The upper and lower graphs are respectively the imaginary parts of $\omega _{1,-1}$ and $\omega _{0}$. 
The vertical and horizontal axes are respectively the relative velocity and wave number of the $x$ direction, which is normalized by the sound 
velocity $c_{s} = \sqrt{c_{0}n_{0}/2M}$ and the healing length $\xi = \hbar /\sqrt{2Mc_{0}n_{0}}$. 
The imaginary parts of $\omega _{1,-1}$ and $\omega _{0}$ have finite values for any relative velocity $V_{R}$. 
Therefore, the initial state is unstable even for $V_{R}=0$. This is in contrast to Fig. 1, which shows the antiferromagnetic case.} 
\end{center}
\end{figure}

We can write the wave functions as
\begin{equation}
\psi _{m} = \psi _{m}^{(0)} +\delta \psi _{m}.
\end{equation}
Our system is homogeneous so that we can express the small deviation $\delta \psi _{m}$ by plane waves as
\begin{equation}
\delta \psi _{m} = ( u_{m} e^{i(\bm{k}\cdot\bm{r}-\omega t)} - v_{m}^{*} e^{-i(\bm{k}\cdot\bm{r}-\omega t)}) e^{-iA_{m}}, 
\end{equation}
\begin{equation}
 \begin{pmatrix} 
A_{1} \\
A_{0}  \\
A_{-1}
\end{pmatrix}
= 
\begin{pmatrix} 
\frac{\mu _{1}}{\hbar}t - \frac{M}{2\hbar} \bm{V_{R}} \cdot \bm{r} \\
\frac{\mu _{0}}{\hbar}t  \\
\frac{\mu _{-1}}{\hbar}t + \frac{M}{2\hbar} \bm{V_{R}} \cdot \bm{r} \\
\end{pmatrix}
,
\end{equation}
where $\mu _{0}$ is $(\mu _{1} + \mu _{-1})/2$. Substituting Eqs. (11)--(13) into Eq. (1) and neglecting the quadratic terms of the small deviation, we obtain the following equations: 
\begin{equation}
 \bm{M}_{0}
\begin{pmatrix} 
u_{0} \\
v_{0} 
\end{pmatrix}
=
\hbar \omega
\begin{pmatrix} 
u_{0} \\
v_{0} 
\end{pmatrix}
,
\end{equation}
\begin{equation}
\bm{M}_{1,-1}
\begin{pmatrix} 
u_{1} \\
v_{1} \\
u_{-1} \\
v_{-1}
\end{pmatrix}
=
\hbar \omega
\begin{pmatrix} 
u_{1} \\
v_{1} \\
u_{-1} \\
v_{-1}
\end{pmatrix}
,
\end{equation}
where
\begin{equation}
\bm{M}_{0} = 
\begin{pmatrix} 
\epsilon _{k} + c_{1}n_{0} -\frac{MV_{R}^{2}}{8} &  -c_{1}n_{0} \\
c_{1}n_{0}  & -\epsilon _{k} - c_{1}n_{0} + \frac{MV_{R}^{2}}{8}\\
\end{pmatrix}
,
\end{equation}

\begin{equation}
\bm{M}_{1,-1} = 
\begin{pmatrix} 
h_{+} & - \frac{n_{0}(c_{0}+c_{1})}{2} & \frac{n_{0}(c_{0}-c_{1})}{2} & -\frac{n_{0}(c_{0}-c_{1})}{2}\\
\frac{n_{0}(c_{0}+c_{1})}{2}  & -h_{-} & \frac{n_{0}(c_{0}-c_{1})}{2} & - \frac{n_{0}(c_{0}-c_{1})}{2}\\
\frac{n_{0}(c_{0}-c_{1})}{2} & - \frac{n_{0}(c_{0}-c_{1})}{2} & h_{-} & -\frac{n_{0}(c_{0}+c_{1})}{2}\\
\frac{n_{0}(c_{0}-c_{1})}{2} & - \frac{n_{0}(c_{0}-c_{1})}{2} & \frac{n_{0}(c_{0}+c_{1})}{2} & -h_{+}
\end{pmatrix}
,
\end{equation}
with $\epsilon _{k} = \hbar ^{2}k^{2}/2M$ and $h_{\pm} = \epsilon _{k} + n_{0}(c_{0}+c_{1})/2 \pm \hbar \bm{V}_{R}\cdot \bm{k} /2$.

It follows that the small deviations of the $m= \pm 1$ components couple with each other, but that the small deviation of the $m=0$ component develops independently of the small deviation of the $m= \pm 1$ components. 
This is because the initial state of Eq. (10) does not have any $m=0$ component. 
Since Eqs. (14) and (15) are eigenvalue problems, we can obtain the dispersion relations:  
\begin{equation}
(\hbar \omega _{0})^{2} = ( \epsilon _{k} + c_{1}n_{0} - \frac{1}{8}M V_{R}^2 )^{2} - c_{1}^{2}n_{0}^{2} 
\end{equation}
\begin{eqnarray}
(\hbar \omega _{1,-1})^{2} = \epsilon _{k}^{2} + (c_{0} + c_{1})n_{0}\epsilon _{k} + \frac{1}{4}(\bm{V}_{R}\cdot \hbar\bm{k})^{2} \nonumber \\ 
\pm \sqrt{(\bm{V}_{R}\cdot \hbar\bm{k})^{2} \epsilon _{k}(\epsilon _{k} + c_{0}n_{0} + c_{1}n_{0}) + n_{0}^{2}(c_{0}-c_{1})^{2} \epsilon _{k}^{2}},
\end{eqnarray}
where $\omega _{0}$ and $\omega _{1,-1}$ are the eigenfrequencies of Eq. (14) and (15), respectively. 

The dispersion relations of Eqs. (18) and (19) have dynamically unstable regions where the imaginary parts of $\omega _{1,-1}$ and $\omega _{0}$ become finite. 
Figures 1 and 2 show the imaginary parts of the dispersion relations for the antiferromagnetic ($c_{0}/c_{1}=20, c_{0} > 0$) and ferromagnetic cases ($c_{0}/c_{1}=-20, c_{0} > 0$), respectively. 
Here, the wave number and the velocity are normalized by the coherence length $\xi = \hbar /\sqrt{2Mc_{0}n_{0}}$ and the sound velocity $c_{s} = \sqrt{c_{0}n_{0}/2M}$. 
Experiments typically use $^{23}$Na and $^{87}$Rb atoms. The interaction parameters for $^{23}$Na atoms satisfy $c_{0}/c_{1} \sim 20$ and $c_{0} > 0$, and 
those for $^{87}$Rb atoms satisfy $c_{0}/c_{1} \sim -200$ and $c_{0} > 0$. 
If we had used $c_{0}/c_{1} \sim -200$ and $c_{0} > 0$, it would take much longer for the instability to occur. In this study, to extract the dynamics characteristic of the ferromagnetic 
interaction, we use $c_{0}/c_{1} = -20$ and $c_{0} > 0$. 

In the following, we explain the character of the dispersion relations and show the kinds of dynamics that they are expected to give, which 
can be confirmed by numerical calculations based on the GP equation (see Sec. IV).

In the antiferromagnetic case, the imaginary part of $\omega _{0}$ has finite values for all relative velocities $V_{R}$ except $V_{R}=0$,
whereas the imaginary part of $\omega _{1,-1}$ has finite values only for relative velocities larger than some critical value $V_{c}$. 
The critical velocity $V_{c}/c_{s}$ of $\omega _{1,-1}$ in Fig. 1 (i.e., the lowest relative velocity for which the imaginary part of $\omega _{1,-1}$ is finite) is 
$2\sqrt{2c_{1}/c_{0}} \sim 0.63$. 
For $0 < V_{R} < V_{c}$, only $\omega _{0}$ has an imaginary part.  This means that the density is modulated only for the $m=0$ component. 
However, we expect that the density will be modulated for the $m= \pm 1$ components too. 
The instability occurs for the $m= \pm 1$ components even though the imaginary part of $\omega _{1,-1}$ in Fig. 1 vanishes
because Eqs. (8) and (9) show that increasing the particle number of the $m=0$ component reduces the particle number of the $m=\pm 1$ components.
Consequently, the instability of the $m=\pm 1$ components can occur. 
When the relative velocity $V_{R}$ exceeds $V_{c}$, the amplitude of the imaginary part of $\omega _{1,-1}$ is larger than that of $\omega _{0}$. 
Thus, the instability of the $m=\pm 1$ components occurs faster than that of the $m=0$ component.

In the ferromagnetic case, the dispersion relations for $\omega _{1,-1}$ and $\omega _{0}$ have finite imaginary parts at any arbitrary relative velocity $V_{R}$. 
This is in contrast to the antiferromagnetic case and it implies that the initial state of Eq. (10) is unstable  even without a counterflow. 
In the case $V_{R}\sim 0$, the instability of all components occurs nearly at the same time because the amplitude of the imaginary parts of 
$\omega _{1,-1}$ is almost same as that of $\omega _{0}$. On the other hand, in the case $V_{R} > 0$, the instability of the $m=\pm 1$ components grows faster than that of the $m=0$ 
component, which reflects the amplitude of the imaginary part of $\omega _{1,-1}$ and $\omega _{0}$. 

Finally, we discuss the isotropy and anisotropy of the dispersion relation about the direction of the relative velocity. Equation (18) shows that $\omega _{0}$ depends not on 
the direction of the relative velocity but on its amplitude. On the other hand, $\omega _{1,-1}$ depends on the direction of the relative velocity as well as its amplitude. 
Thus, the dynamics obtained by the GP equation exhibits different behaviors depending on which eigenfrequencies give larger imaginary parts. 
If $\omega _{0}$ ($\omega _{1,-1}$) is dominant, the early dynamics will be isotropic (anisotropic).

These results obtained from the dispersion relation are consistent with numerical calculations based on the GP equation (see Sec. IV).

\begin{figure}[t]
\begin{center}
\includegraphics[keepaspectratio, width=9cm,clip]{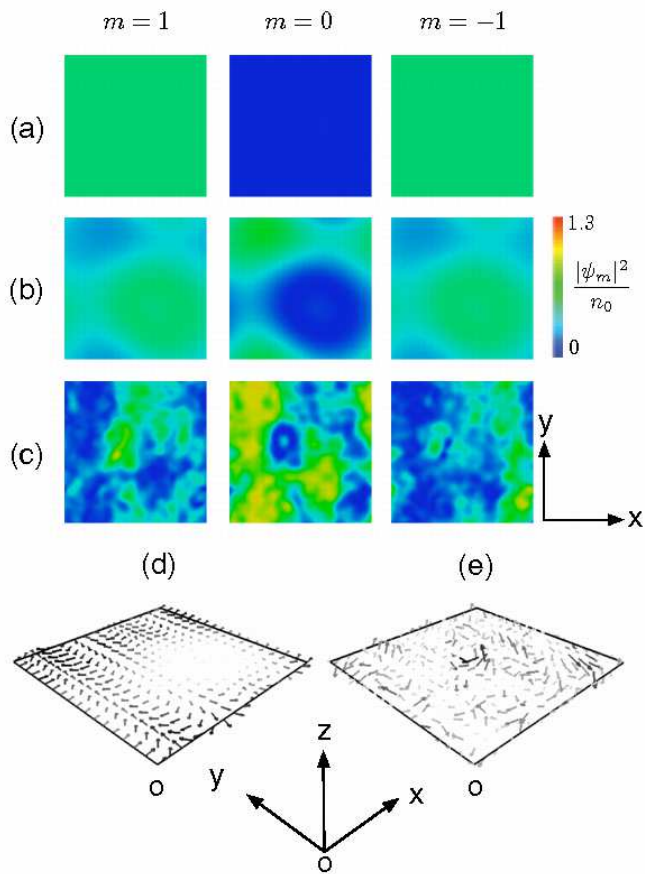}
\caption{Density profiles of the $m = 1, 0, -1$ components for the antiferromagnetic case at $t/\tau =$ (a) $0$, (b) $315$, and (c) $1500$.
(d) and (e) profiles of the spin density vector corresponding to (b) and (c), respectively. The field of view of each image is $128\xi \times 128\xi$. 
The shading of the arrows in (d) and (e) denote the magnitude of the spin density vector. 
These results are for numerical calculations with $c_{0}/c_{1} = 20$, $c_{0} > 0$, and $V_{R}/c_{s} = 0.39$. } 
\end{center}
\end{figure}

\section{Counterflow instability}

We investigate the dynamics of a spin-1 spinor BEC induced by counterflow by performing numerical calculations based on the GP equation. 
The dynamics is mainly classified according to whether the spin-dependent interaction is antiferromagnetic or ferromagnetic. In this section, we present the
detailed dynamics for both cases. 

\subsection{Antiferromagnetic interaction case}
The dynamics for the antiferromagnetic interaction induced by counterflow instability are shown. 
The dynamics strongly depends on whether $0 < V_{R} < V_{c}$ or $V_{c} < V_{R}$.
As shown in Sec. III, in the former case, the instabilities for 
all three components are expected to grow simultaneously, whereas the instability for the $m= \pm 1$ components is expected to grow faster than that for the $m=0$ component in the latter case. 

\begin{figure}[t]
\begin{center}
\includegraphics[keepaspectratio, width=9cm,clip]{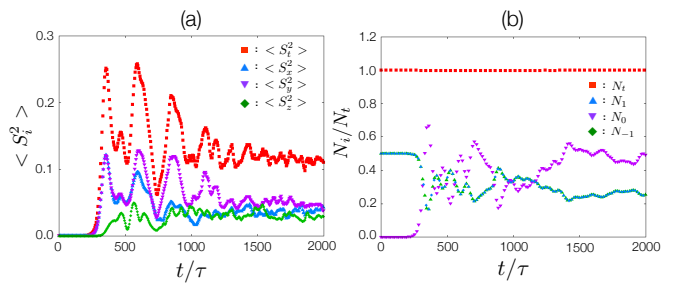}
\caption{Time dependence of the quantities $<S^{2}_{i}>$ ($i = t, x, y, z$) and the particle number $N_{i}$ ($i = t, x, y, z$) for the antiferromagnetic case. 
$S^{2}_{i}$ ($i = t, x, y, z$) is defined by Eqs.(20) and (21).  
These results are for numerical calculations with $c_{0}/c_{1} = 20$, $c_{0} > 0$, and $V_{R}/c_{s} = 0.39$.} 
\end{center}
\end{figure}

First, we present the dynamics for the case $0 < V_{R} < V_{c}$. 
Figure 3 shows the density profiles of all components obtained by numerical calculations with $V_{R}/c_{s} = 0.39$. 
In this case, the critical velocity $V_{c}/c_{s}$ is about $0.63$. 
Figure 3(a) shows the initial state in which the $m= \pm 1$ components have a velocity $\bm{V}_{R}$ relative to the counterflow. 
Figure 3(b) shows the density profiles at $t/\tau = 315$, where instabilities occur in all components. 
The instability in Fig. 3(b) appears as an isotropic density modulation independent of the direction of the relative velocity $\bm{V}_{R}$: the observed 
density modulation is circular. These results can be understood by considering the imaginary part of the dispersion relation of Eqs. (18) and (19). 
For $0 < V_{R} < V_{c}$, the instability is induced by the imaginary part of the dispersion relation of the $m=0$ component, which is independent of the direction of 
the relative velocity $V_{R}$. In addition, as pointed out in Sec. III, the instability of the $m=0$ component causes the instability of the $m=\pm 1$ components through Eq. (8). 
Thus, the instability of the $m= \pm 1$ components is caused by isotropic density modulation of the $m=0$ component and thus it also exhibits isotropic modulation, as shown in Fig. 3(b).
Hence, all components exhibit circular density modulation. 
After a certain time, the density modulation of all components becomes very complicated, as shown in Fig. 3(c).
After Fig. 3(b), the circular modulation expands in every component. This results in a narrow path in the low density region. 
Thus, in the transition from Fig. 3(b) to (c), density modulation with various wave numbers grows with increasing time. 
As a result, the circular density modulation in Fig. 3(b) disappears. 

\begin{figure}[t]
\begin{center}
\includegraphics[keepaspectratio, width=9cm,clip]{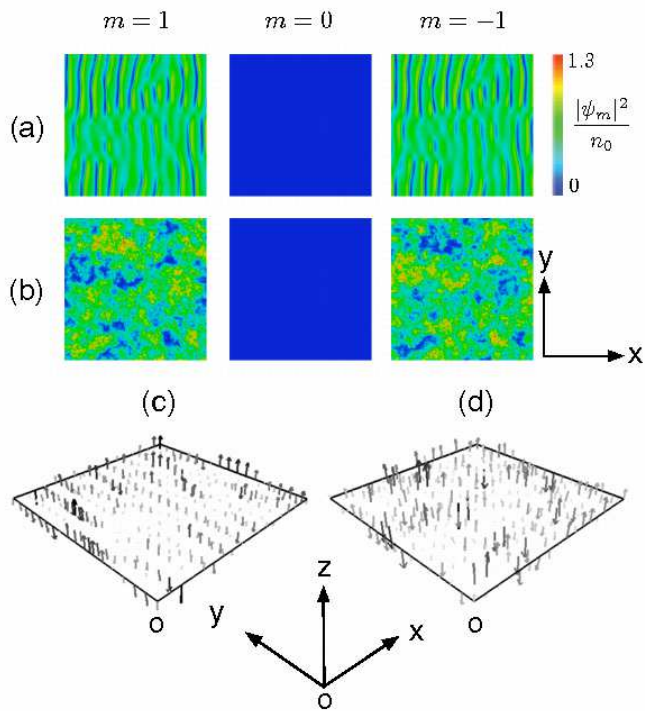}
\caption{Density profiles of the $m = 1, 0, -1$ components for the antiferromagnetic case at (a) $t/\tau =$ (a) $70$ and (b) $1000$. 
(c) and (d) Profiles of the spin density vector corresponding to (a) and (b), respectively. The field of view of each image is $128\xi \times 128\xi$.
The shading of the arrows in (c) and (d) denotes the magnitude of the spin density vector. 
These results for numerical calculations with $c_{0}/c_{1} = 20$, $c_{0} > 0$, and $V_{R}/c_{s} = 1.57$.} 
\end{center}
\end{figure}

\begin{figure}[b]
\begin{center}
\includegraphics[keepaspectratio, width=9cm,clip]{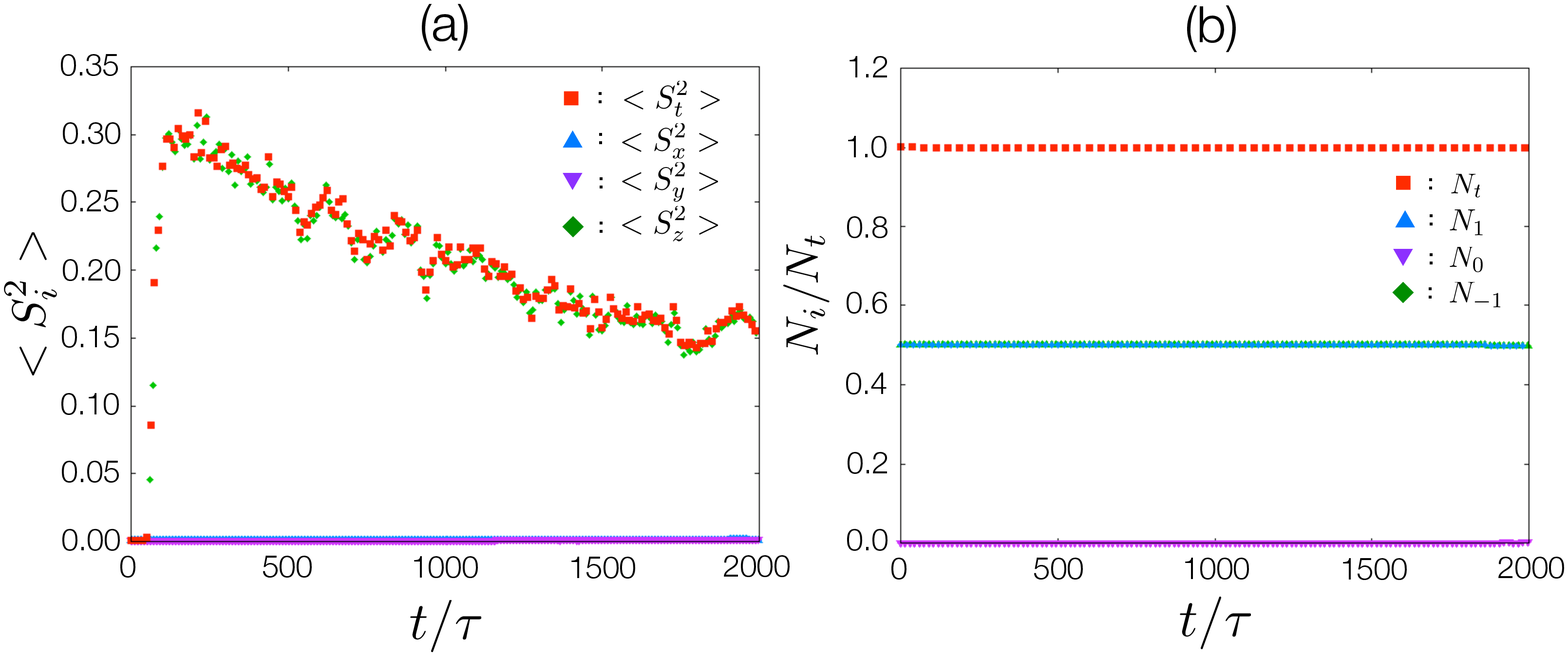}
\caption{Time dependence of quantities $<S^{2}_{i}>$ ($i = t, x, y, z$) and particle number $N_{i}$ ($i = t, x, y, z$) for the antiferromagnetic case. 
$S^{2}_{i}$ ($i = t, x, y, z$) is defined by Eqs. (20) and (21). 
These results for numerical calculations with $c_{0}/c_{1} = 20$, $c_{0} > 0$, and $V_{R}/c_{s} = 1.57$.} 
\end{center}
\end{figure}

The behavior of the spin density vector is shown in Figs. 3(d) and (e), which correspond to the density profile in Figs. 3(b) and (c), respectively. 
The spin density vector in Fig. 3(d) almost lies in the $x-y$ plane. In the early stages of the instability at $t/\tau = 315$, the density modulation of the $m= 1$ component
almost overlaps that of the $m=-1$ component.
Thus, the $z$ component of the spin density vector is very small. 
As time progresses, the modulation of high wave numbers increases, and the $m= \pm 1$ components become less overlapped.
Therefore, the $z$ component grows as shown in Fig. 3(e).

To understand the time dependence of the magnitude of the spin density vector, we numerically calculate the following quantities:
\begin{equation}
<{S^{2}_{i}}> = \frac{1}{n_{0}^{2}A} \int s_{i}(\bm{r})^{2} d\bm{r} \hspace{1cm} (i = x, y, z),
\end{equation}
\begin{equation}
<{S^{2}_{t}}> = \sum _{i=x,y,z} <{S_{i}^{2}}>, 
\end{equation}
where $A$ is the area of the system.
Their time dependence is shown in Fig. 4(a). The result of $<{S^{2}_{i}}>$ means that the instability of all components occurs at $t/\tau \sim 300$ because growth of 
$<{S^{2}_{x}}>$ and $<{S^{2}_{y}}>$ requires the instability of $m=0$ component and that of $<{S^{2}_{z}}>$ requires the instability of the $m= \pm 1$ components. 
Figure 4(b) shows that the particle number of the $m=0$ component increases rapidly at $t/\tau \sim 300$, which corresponds 
to the occurrence of density modulation in Fig. 3(b). 
Hence the instability induced by the counterflow starts to exchange the particle number  among the three components and causes the spin density vector to increase.

Next, we present the dynamics for $V_{c} < V_{R}$. 
Figure 5 shows the density profile for $V_{R}/c_{s} = 1.57$. 
Figures 5(a) shows that the density modulation of the $m=\pm 1$ components is much greater than that of the $m=0$ component because 
${\rm{Im}}[\omega _{1,-1}]$ is larger than ${\rm{Im}}[\omega _{0}]$ in Fig. 1.
This density profile differs from that for $0 < V_{R} < V_{c}$ because the density modulation in Fig. 5(a) is anisotropic. 
This is understood by the dispersion relation for $\omega _{1,-1}$ in Eq. (19) which depends on the direction of the relative velocity $\bm{V}_{R}$. 
The low density region of the $m=\pm 1$ components in Fig. 5(a) is nucleated due to the growth of the density stripe perpendicular to the relative velocity $\bm{V}_{R}$. 
The interval of the stripe  is consistent with the most unstable wave number obtained for ${\rm{Im}}[\omega _{1,-1}]$ in Fig. 1.
Through this stripe in the low density region in Fig. 5(a), the phase of each wave function rapidly changes by about $\pi$, whose structure is similar to solitons in one-component BECs. This soliton-like structure soon collapses and the density modulation of the $m=\pm 1$ components become complicated, as shown in Fig. 5(b). 
Through this collapse, the density modulation of the $m = \pm 1$ components becomes isotropic.
Similar dynamics to that shown in Figs. 5(a) and (b) has been reported in two-component BECs \cite{Ishino10, Ishino11, Hamner11, Hoefer11}.
Even after a long time, the density modulation of the $m=0$ component does not increase. However, this result depends on the initial noise (see Sec. IV C).

The spin density vector behaves very differently from the case when $0 < V_{R} < V_{c}$, as shown in Figs. 5(c) and (d). 
In Fig. 5(c), density modulation occurs only in the $m=\pm 1$ components (Fig. 5 (a)), so that the spin density vector is oriented in only $\pm z$ directions. 
In addition, the vector in Fig. 5(c) exhibits the stripe structure as that in the density profiles in Fig. 5(a).
As time passes, the structure in Fig. 5(c) collapses because the density profiles lose the stripe structure as shown in Fig. 5(b).
Then, as the density modulation of the $m=0$ component does not increase, the spin density vector cannot lean in the $x-y$ plane (Fig. 5(d)).
Therefore, domains in which the spin density vector points in the $+z$ or $-z$ direction are formed through the instability and they move around. 

The behaviors of the magnitude of the spin density vector and the particle number of each component for $V_{c} < V_{R}$ differ greatly from those for $0< V_{R} < V_{c}$ too. 
Figure 6 shows the time dependence of $<S^{2}_{i}>$ $(i=t, x, y, z)$ and the particle number of each component.
$<S^{2}_{z}>$ grows rapidly at $t/\tau = 70$, but the particle number of each component remains almost the same as that of the initial state.
This means that the instability of the $m=\pm 1$ components occurs, but that of the $m=0$ does not, which is consistent with the density profiles in Figs. 5(a) and (b). 
This dynamics is almost the same as the behavior of the two-component BEC \cite{Ishino10, Ishino11, Hamner11, Hoefer11}.

\begin{figure}[t]
\begin{center}
\includegraphics[keepaspectratio, width=9cm,clip]{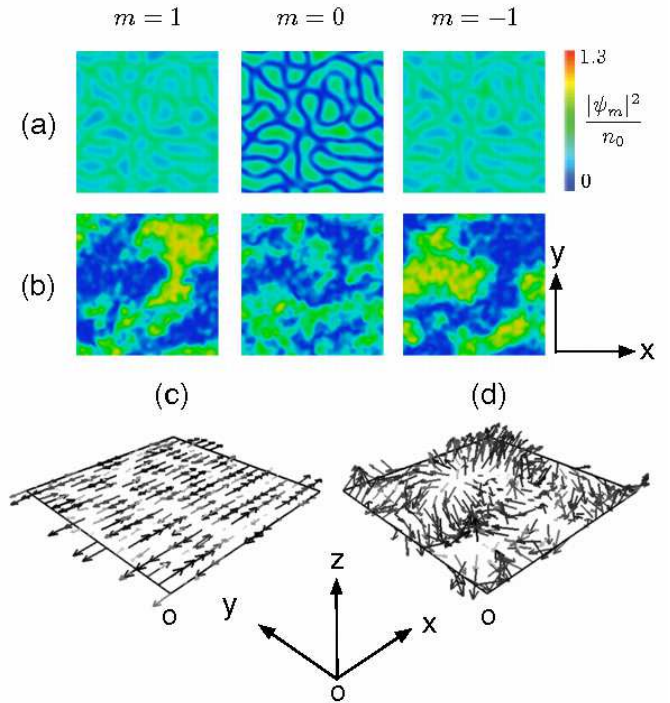}
\caption{Density profiles of the $m = 1, 0, -1$ components in the ferromagnetic case at $t/\tau = $ (a) $175$ and (b) $1000$. 
(c) and (d) profiles of the spin density vector corresponding to (a) and (b), respectively. The field of view of each image is $128\xi \times 128\xi$.
The shading of the arrows in (c) and (d) indicates the magnitude of the spin density vector. 
These results are for numerical calculation with $c_{0}/c_{1} = -200$, $c_{0} > 0$, and $V_{R}/c_{s} = 0$}. 
\end{center}
\end{figure}

\begin{figure}[t]
\begin{center}
\includegraphics[keepaspectratio, width=9cm,clip]{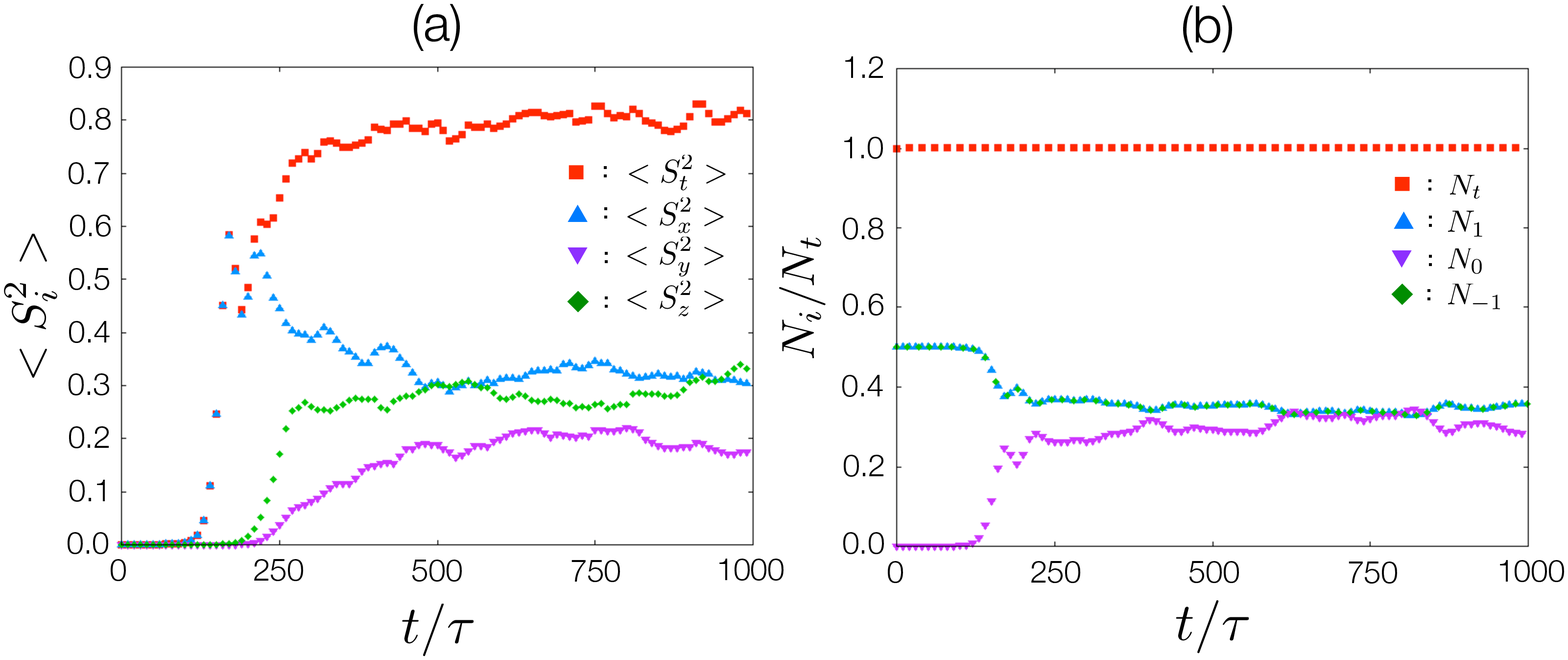}
\caption{Time dependence of quantities $S^{2}_{i}$ ($i = t, x, y, z$) and the particle number $N_{i}$ ($i = t, x, y, z$) for the ferromagnetic case. 
$<S^{2}_{i}>$ ($i = t, x, y, z$) is defined by Eqs. (20) and (21). 
These results are for numerical calculation with $c_{0}/c_{1} = -200$, $c_{0} > 0$, and $V_{R}/c_{s} = 0$.} 
\end{center}
\end{figure}

\subsection{Ferromagnetic interaction case}
This subsection presents the dynamics with the ferromagnetic interaction induced by the counterflow instability. In this case, the imaginary parts of $\omega _{1,-1}$ and $\omega _{0}$
in Fig. 2 do not exhibit a critical velocity, unlike the case for the antiferromagnetic interaction. Hence, the dynamics greatly depends on whether 
the relative velocity is $V_{R} \sim 0$ or $0 < V_{R}$. As shown in Sec. III, in the former case, the instability of all components grows, whereas in the latter case the instability 
of the $m=\pm 1$ components grows faster than that of the $m=0$ component.

We present the dynamics for the case $V_{R} = 0$. 
In this case, there is no counterflow, but instability occurs because the initial state is unstable even for $V_{R} = 0$. 
Figure 7 shows the density profiles of each component obtained by numerical calculations with $V_{R} = 0$. 
As expected, the density profiles do not exhibit anisotropy (Fig. 7(a)) since the initial state with $V_{R} = 0$ is isotropic.
Isotropic density modulation occurs; it corresponds to the most unstable wave number in Fig. 2. 
The instability grows considerably, as shown in Fig. 7(b). 

The spin density vector corresponding to the density profiles in Figs. 7(a) and (b) are shown in Figs. 7(c) and (d), respectively. 
When the instability occurs at $t/\tau \sim 175$, the density modulation of the $m= 1$ component overlaps with that of the $m=-1$ component.
Thus, the spin density vector almost lies in the $x-y$ plane, as shown in Fig. 7(c). 
With increasing time, the overlap decreases, so that the vector points in various directions, as shown in Fig. 7(d). 
This behavior of the spin density vector is similar to that for the antiferromagnetic case with $0<V_{R}<V_{c}$ (see Figs. 3(d) and (e)).
Note that the magnitude of the vector in the ferromagnetic case is larger than that in the antiferromagnetic 
case (indicated by the shading of the arrows in the vector plots); this is discussed later.

Figure 8 shows the time dependences of $<S^{2}_{i}>$ $(i=t, x, y, z)$ and the particle number. 
These results show that the instabilities for all components occur almost simultaneously. 
The density profile for each component in Fig. 7 is consistent with these results. 

We present the dynamics for the case $0 < V_{R}$. 
Figures 9 and 10 show numerical results for $V_{R} /c_{s} = 1.96$. 
Figure 9(a) shows the instability of the $m= \pm 1$ components; density modulation of the $m=0$ component does not occur because ${\rm{Im}}[\omega _{1,-1}] > {\rm{Im}}[\omega _{0}]$. 
The low density regions of the $m=\pm 1$ components in Fig. 9(a) are the soliton-like structure, which collapse (Fig. 9(b)). 
As time progresses, the instability of the $m=0$ component develops, as shown in Fig. 9(c). 
The spin density vector for this case is shown in Figs. 9(d) and (e), which correspond to Figs. 9(a) and (c), respectively. 
In the early stages of the instability, density modulation occurs only in the $m=\pm 1$ components, so that 
the spin density vector points in the $\pm z$ directions, which is similar to Fig. 5(c). As time increases, the $m=0$ component grows. 
Thus, the $x$ and $y$ components of the spin density vector become large so that the spin density vector points in various directions, as shown in Fig. 9(e). 
These behaviors of the vector are consistent with the time dependences of the $<S_{i}^{2}>$ and the particle number of each component shown in Fig. 10.

These results reveal that there are obvious differences in the behaviors of the magnitude of the spin density vector for the antiferromagnetic and ferromagnetic cases.
Figures 4, 6, 8, and 10 show that in the antiferromagnetic case $<S^{2}_{t}>$ tends to decrease, whereas it tends to increase in the ferromagnetic case. 
This gives rise to the different behaviors of the PDF of the magnitude of the spin density vector, which is very important for the turbulence 
of a spin-1 spinor BEC. The details are described in Sec. V. 

\begin{figure}[t]
\begin{center}
\includegraphics[keepaspectratio, width=9cm,clip]{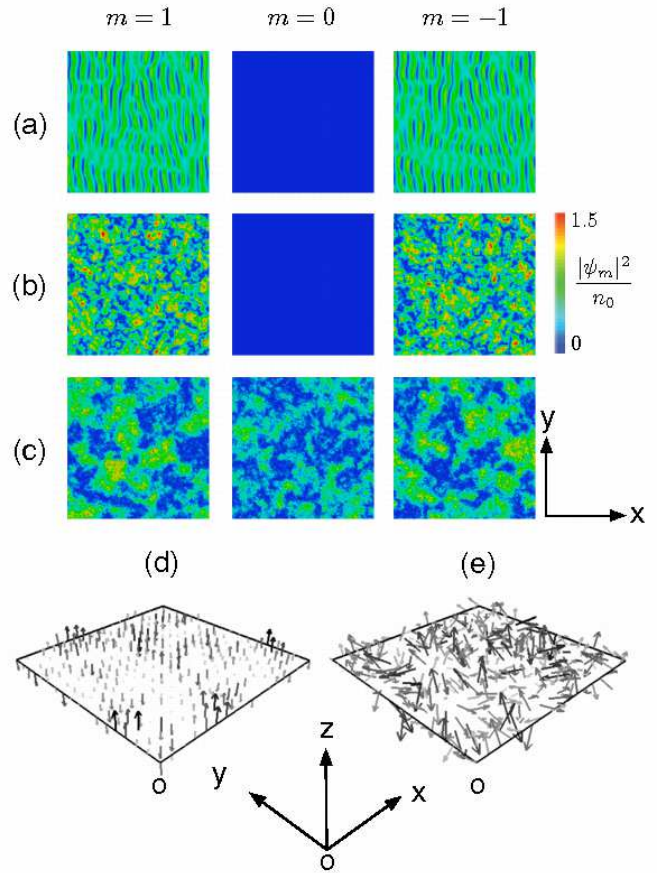}
\caption{Density profiles of the $m = 1, 0, -1$ components for the ferromagnetic case at $t/\tau =$ (a) $40$, (b) $125$, and (c) $500$. 
(d) and (e) profiles of the spin density vector corresponding to (a) and (c), respectively. The field of view of each image is $128\xi \times 128\xi$.
The shaping of the arrows in (d) and (e) indicates the magnitude of the spin density vector. 
These results are for numerical calculations with $c_{0}/c_{1} = -200$, $c_{0} > 0$, and $V_{R}/c_{s} = 1.96$.} 
\end{center}
\end{figure}

\begin{figure}[t]
\begin{center}
\includegraphics[keepaspectratio, width=9cm,clip]{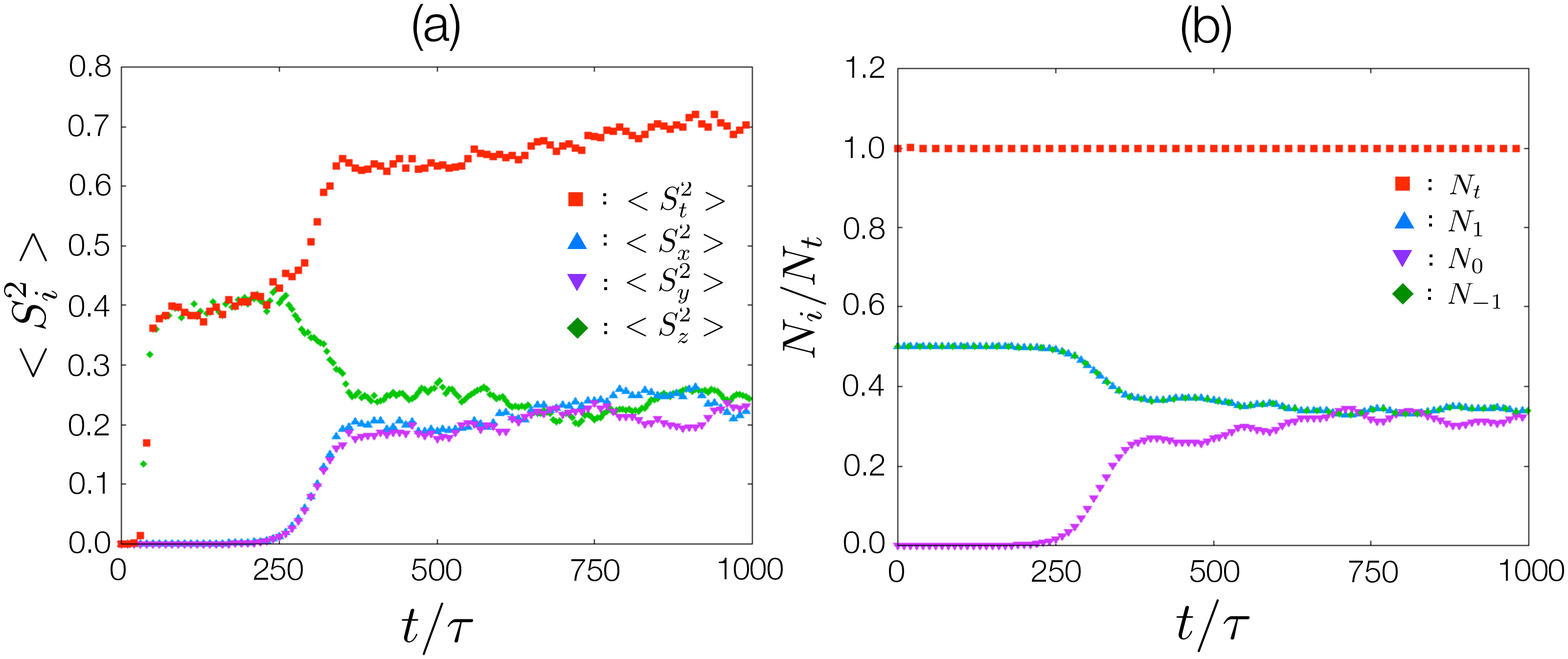}
\caption{Time dependences of the quantities $<S^{2}_{i}>$ ($i = t, x, y, z$) and the particle number $N_{i}$ ($i = t, x, y, z$) for the ferromagnetic case. 
$S^{2}_{i}$ ($i = t, x, y, z$) is defined by Eqs. (20) and (21). 
These results are for numerical calculations with $c_{0}/c_{1} = -200$, $c_{0} > 0$, and $V_{R}/c_{s} = 1.96$.} 
\end{center}
\end{figure}

\subsection{Dependence of the dynamics on the initial noise}
We discuss the dependence of the dynamics on the initial noise. 
In our numerical calculations, small white noise is added to the initial state. Different samples of the white noise are available, although the magnitude is fixed. 
Our numerical results can be classified into four categories: (I) antiferromagnetic case with a small relative velocity (Fig. 3); (II) antiferromagnetic case with a large relative velocity (Fig. 5); (III) ferromagnetic case with a small relative velocity (Fig. 7); (IV) ferromagnetic case with a large relative velocity (Fig. 9).
We performed some numerical calculations and we empirically observed a strong dependence on the noise sample only for (II).
This strong dependence is related to the growth of the $m=0$ component. In this case, the instability of the $m=\pm 1$ components occurs first (Figs. 5 and 6), which is 
independent of the noise. However, numerical calculations reveal that as time increases, the $m=0$ component may grow depending on the noise sample used. 
We cannot control whether growth occurs or not.
In this paper, we consider the results that do not depend on the initial noise; thus, Figs. 5 and 6 show only the results obtained before the $m=0$ component grows. 

For cases (I), (III), and (IV), the dynamics is qualitatively unchanged even when the initial noise sample is varied.

\section{Turbulence in a spin-1 spinor BEC}

We find that, in the ferromagnetic case, the spectrum of the spin-dependent interaction energy in the turbulent state obeys the $-7/3$ power law, whereas that in the antiferromagnetic case does not. 
This result was obtained by numerical calculations and it can be understood in terms of scaling analysis. 
This section mainly considers the spectrum of the spin-dependent interaction energy and its time dependence.

The methodology usually used for analyzing classical turbulence is applied to a spin-1 spinor BECs in this study. 
Here, we briefly review turbulence in classical fluids \cite{Frisch,K41}. 
Many studies focus on statistical quantities and laws because they reflect properties characteristic of the complicated motion in turbulence. 
One quantity often investigated is the kinetic energy spectrum. 
Kolmogorov proposed that the kinetic energy spectrum of incompressible fluid obeys a
$-5/3$ power law in fully developed homogeneous isotropic turbulence; this can be demonstrated by making several assumptions. 
One assumption is that the kinetic energy flux is independent of the wave number. This means that the kinetic energy 
is transported to a high wave number with a constant energy flux in the wave number region that obeys  the $-5/3$ power law.
This result has been confirmed by many numerical calculations and experiments. 
The inertial term in the Navier--Stokes equation was found to play a dominant role in energy transfer. 
These results show the self-similarity of the velocity field in wave number space. 
On the other hand, in real space, a Richardson cascade is believed to occur in which large vortices become smaller through reconnections of vortices. 
However, this has not been confirmed yet.

We focus on the spectrum of the spin-dependent interaction energy in the turbulence of a spin-1 spinor BEC. 
In analogy with classical turbulence, the flux of the spin-dependent interaction energy is expected to be independent of 
the wave number, which gives rise to the spectrum characteristics of this system. 

We derive an expression for the spectrum of the spin-dependent interaction energy.
The spin-dependent interaction energy $E_{s}$ per unit area is given by 
\begin{equation}
E_{s} = \frac{c_{1}}{2A} \int \bm{s}(\bm{r})^{2} d\bm{r}.
\end{equation}
We expand the spin density vector $\bm{s}(\bm{r})$ with plane waves:
\begin{equation}
\bm{s}(\bm{r}) = \sum _{\bm{k}} \tilde{\bm{s}}(\bm{k}) e^{i\bm{k}\cdot\bm{r}}.
\end{equation}
The spin-dependent interaction energy $E_{s}$ is represented by $\tilde{\bm{s}}(\bm{k})$ as
\begin{equation}
E_{s} = \frac{c_{1}}{2} \sum _{\bm{k}} |\tilde{\bm{s}}(\bm{k})|^{2}. 
\end{equation}
Therefore, the energy spectrum of spin-dependent interaction energy is given by
\begin{equation}
E_{s} (k) = \frac{c_{1}}{2 \Delta k} \sum _{k<|\bm{k}_{1}|<k+\Delta k} |\tilde{\bm{s}}(\bm{k}_{1})|^{2},  
\end{equation}
where $\Delta k$ is $2\pi /L$ for a system size $L$.

\subsection{Ferromagnetic interaction case}

Our numerical results reveal that the spectrum of the spin-dependent interaction energy in the ferromagnetic case obeys the power law shown in Fig. 11 (a).
This spectrum is numerically calculated at $t/\tau = 5000$, which is a sufficiently long time after the occurrence of the instability. 
This is found by the time dependence of $<S^{2}_{i}>$ ($i = t, x, y, z$) shown in Fig. 11 (b). 
The spectrum in Fig. 11 (a) has two regions that are separated by the wave number $2\pi /\xi _{s}$, which corresponds to the spin coherence length $\xi _{s} = \hbar / \sqrt{2M|c_{1}|n_{0}}$, (i.e., 
the characteristic scale of spin structures such as domain walls and polar core vortex \cite{Saito07}). 
In the low wave number region ($k < 2\pi /\xi _{s}$), the spectrum obeys the power law, whereas it does not in the high wave number region ($2\pi /\xi _{s} < k $). 
The origin of this power law in the low wave number region is discussed below. 

\begin{figure}[t]
\begin{center}
\includegraphics[keepaspectratio, width=9cm,clip]{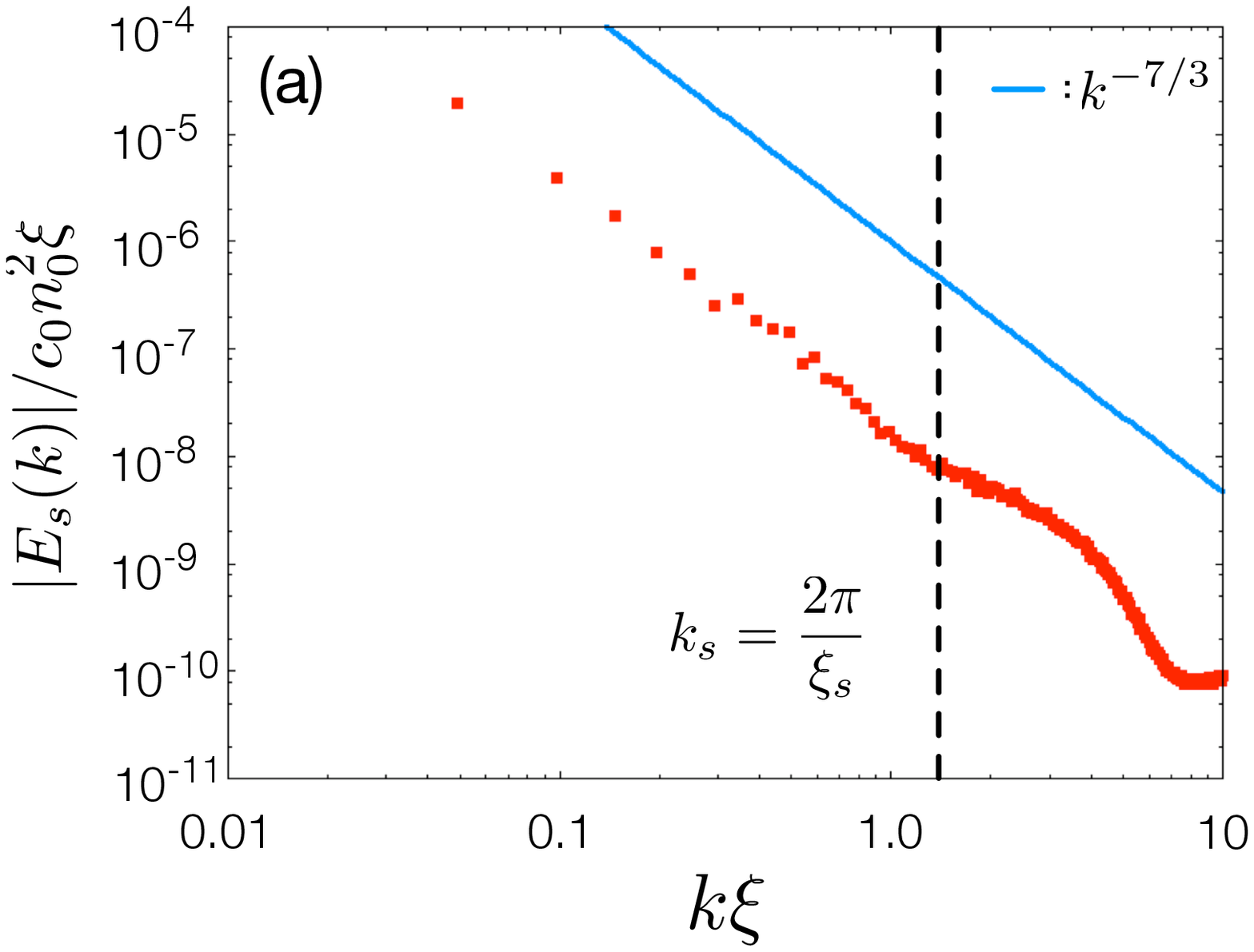}
\includegraphics[keepaspectratio, width=9cm,clip]{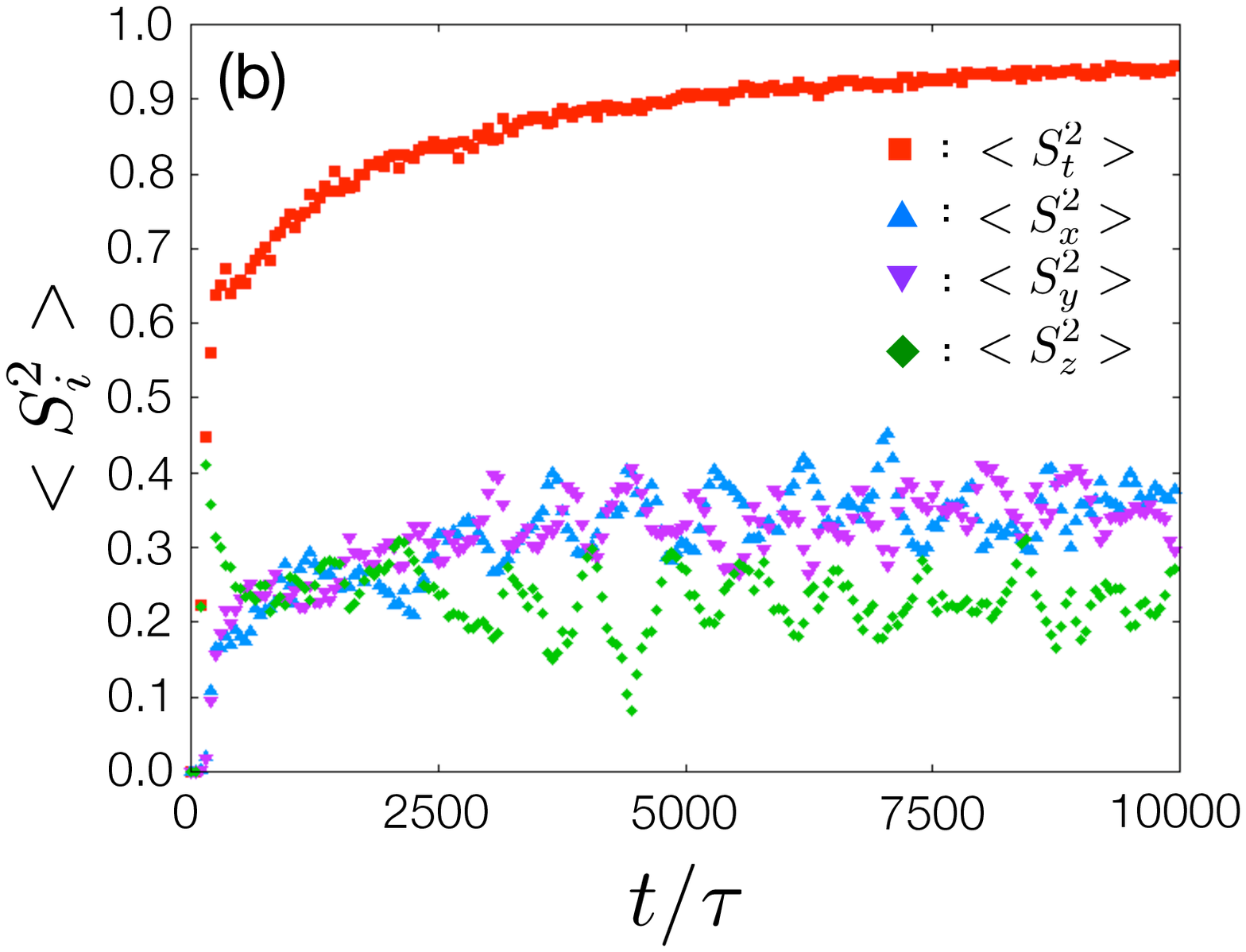}
\caption{Spectrum of the spin-dependent interaction energy $E_{s}(k)$ (upper) and time dependences of the quantities $<S^{2}_{i}>$ ($i = t, x, y, z$) (lower) for the ferromagnetic case at $t/\tau = 5000$. 
In the graph of the spectrum, the red squares and blue solid and black dashed lines show the numerical results for the spectrum, a line proportional to $k^{-7/3}$, and 
the boundary of the wave number corresponding to the spin coherent length, respectively. 
This spectrum is obtained by numerical calculations with $c_{0}/c_{1} = -20$, $c_{0} > 0$, $V_{R}/c_{s} = 0.78$. } 
\end{center}
\end{figure}

\begin{figure}[b]
\begin{center}
\includegraphics[keepaspectratio, width=9cm,clip]{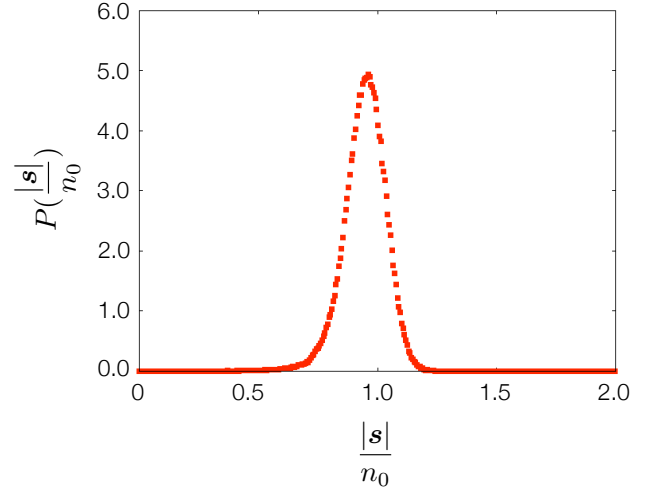}
\caption{Probability density function of the magnitude of the spin density vector for the ferromagnetic case at $t/\tau = 5000$.
This was obtained by performing numerical calculations with the following parameters: $c_{0}/c_{1} = -20$, $c_{0} > 0$, $V_{R}/c_{s} = 0.78$.} 
\end{center}
\end{figure}

We will obtain the power exponent of the spectrum of the spin-dependent interaction energy in the low wave number region ($k < 2\pi /\xi _{s}$) using the GP equation. 
The calculation used to obtain Fig. 11 (a) was for the ferromagnetic case, where the magnitude of the spin density vector is expected 
to have some large value in their PDF. 
To confirm this, we calculated the PDF of the magnitude of the spin density vector. 
The PDF in Fig. 12 has a sharp peak, indicating that the magnitude of the spin density vector tends to be $n_{0}$ in the turbulent state.
This observation allows us to write the macroscopic wave function in the turbulent state with the ferromagnetic interaction as
\begin{equation}
\begin{pmatrix} 
\psi _{1} \\
\psi _{0} \\
\psi _{-1}
\end{pmatrix}
\sim \sqrt{n_{0}} {\rm{e}}^{i(\phi - \gamma)}
\begin{pmatrix} 
{\rm{e}}^{-i\alpha} \cos ^{2}{\frac{\beta}{2}} \\
\frac{1}{\sqrt{2}} \sin{\beta} \\
{\rm{e}}^{i \alpha} \sin ^{2}{\frac{\beta}{2}}
\end{pmatrix}
,
\end{equation}
where $\alpha$, $\beta$, and $\gamma$ are the Euler angles in spin space and $n _{0}$ is the total density of the initial state. 
This wave function leads to
\begin{equation}
\bm{s} = n_{0} ( \sin{\beta}\cos{\alpha}, \sin{\beta}\sin{\alpha}, \cos{\beta} ). 
\end{equation} 
In addition, the total density is assumed to be time-independent because $c_{0} \gg c_{1}$. 
In this condition, the spin-independent interaction energy is larger than the kinetic and spin-dependent interaction enegies, so that 
the total density $n$ has a weak time dependence. 
We can obtain the time evolution equation of $\hat{\bm{s}} = \bm{s} /n_{0}$ from Eqs. (1) and (26):
\begin{equation}
\frac{\partial}{\partial t} \hat{\bm{s}} + (\bm{v} \cdot \bm{\nabla}) \hat{\bm{s}} = \frac{\hbar}{2M} \hat{\bm{s}} \times [ \nabla ^{2} \hat{\bm{s}} + (\bm{a} \cdot \bm{\nabla}) \hat{\bm{s}} ], 
\end{equation}
\begin{equation}
\bm{v} = \frac{\hbar}{2Mn_{0}i} \sum _{m=-1} ^{1} ( \psi _{m}^{*} \bm{\nabla} \psi _{m} - \psi _{m} \bm{\nabla} \psi _{m}^{*} )
\end{equation}
with $\bm{a} = (\nabla n _{0})/n_{0}$ \cite{Lamacraft08,Kudo10, Kudo11}. 
In our case, the total density is approximately $n_{0}$ in the initial state of Eq. (10), so that $\bm{a}$ vanishes. 
Therefore, the following equation can be used to calculate the power exponent of the spectrum:
\begin{equation}
\frac{\partial}{\partial t^{'}} \hat{\bm{s}} + (\frac{\bm{v}}{c_{s}} \cdot \bm{\nabla} ^{'}) \hat{\bm{s}} =  \hat{\bm{s}} \times \bm{\nabla} ^{' 2} \hat{\bm{s}},  
\end{equation}
where space and time are respectively normalized by $\xi$ and $\tau$ ($t^{'} = t/\tau$, $\bm{\nabla}^{'} =  \xi \bm{\nabla}$). 
In the early stages of the instability, the wave function has vortices and a soliton-like structure. 
The velocity $\bm{v}$ can be large in their neighborhoods. 
This is similar to the vicinity of vortices in one-component BECs, where the velocity is larger than the sound velocity in the vortex core region whose size is equal to the coherence length. 
However, the number of such defects tends to decrease with time. 
Therefore, the velocity $\bm{v}$ is lower than the sound velocity $c_{s}$ almost everywhere after a sufficiently long time after the instability occurs. 
We confirm that the PDF of the magnitude of the velocity has a peak at about 1/10th of the sound velocity. 
We thus expect that the term on the right-hand side of Eq. (30) is important for transporting the spin-dependent interaction energy to a higher wave number. 
This is different from classical turbulence for which the inertial term in the Navier--Stokes equation is dominant for energy transfer. 

\begin{figure}[t]
\begin{center}
\includegraphics[keepaspectratio, width=9cm,clip]{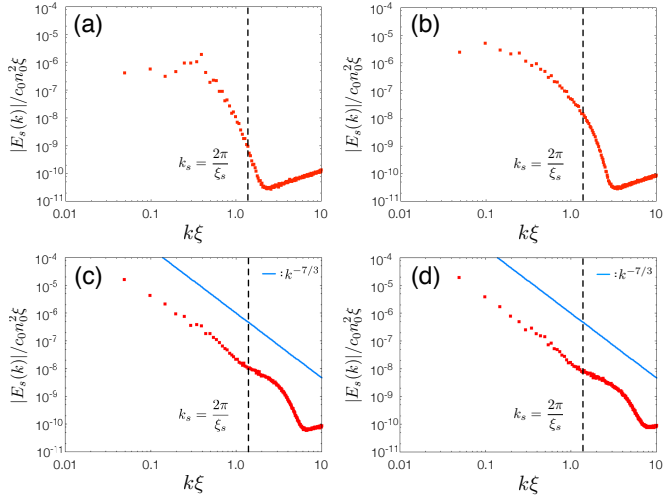}
\caption{Time dependence of spectrum of spin-dependent interaction energy $E_{s}(k)$ for the ferromagnetic case at $t/\tau =$ (a) $125$, (b) $600$, (c) $3000$, and (d) $5000$. 
In the spectrum, the red squares, blue solids, and black dashed lines indicate the numerical results for the spectrum, the line proportional to $k^{-7/3}$, and 
the boundary of the wave number corresponding to the spin coherent length, respectively. 
This was obtained by performing numerical calculations with $c_{0}/c_{1} = -20$, $c_{0} > 0$, $V_{R}/c_{s} = 0.78$.} 
\end{center}
\end{figure}

To understand the behavior of the spectrum in Fig. 11, we consider spin density vector in wave number space. 
We can express Eq. (30) by neglecting the second term on the left-hand side by using $\hat{\tilde{\bm{s}}} = \tilde{\bm{s}}/n_{0}$:
\begin{equation}
\frac{\partial}{\partial t} \hat{\tilde{\bm{s}}} (\bm{k}) = - \sum _{\bm{k}_{1}, \bm{k}_{2}} k_{2}^{2} \hat{\tilde{\bm{s}}} (\bm{k}_{1}) \times \hat{\tilde{\bm{s}}} (\bm{k}_{2}) 
\delta _{\bm{k}, \bm{k}_{1} + \bm{k}_{2}}. 
\end{equation} 
In the following, we apply Kolmogorov-type dimensional analysis to Eq. (31) \cite{Ottaviani91,Watanabe97} under two assumptions:
Equation (31) is invariant under the scale transformation and the flux of the spin-dependent interaction energy 
is independent of the wave number.
We perform the scale transformation $\bm{k} \rightarrow \zeta \bm{k}$, $\bm{t} \rightarrow \eta \bm{t}$ in Eq. (31). 
Then, if $\hat{\tilde{\bm{s}}}$ is transformed to $\zeta ^{-2} \eta ^{-1}\hat{\tilde{\bm{s}}}$, Eq. (31) will be invariant. 
We can then write the dependence $\hat{\tilde{\bm{s}}}$ on $k$ and $t$ as
\begin{equation}
\hat{\tilde{\bm{s}}} \sim k^{-2} t^{-1}. 
\end{equation}
We assume that the flux of the spin-dependent interaction energy is independent of the wave number, which is equivalent to assuming the existence of a region in which the energy is constantly transported. 
This result can be expressed by 
\begin{equation}
\epsilon \sim \frac{\hat{\tilde{\bm{s}}}^{2}}{t_{s}} \sim k^{-4}t_{s}^{-3}, 
\end{equation}
where $\epsilon$ and $t_{s}$ are respectively the energy flux and the characteristic time. 
Using Eqs. (25), (32), and (33), we obtain the $-7/3$ power exponent:
\begin{equation}
E_{s}(k) \sim k^{-1}(k^{-2}t_{s}^{-1})^{2} \sim \epsilon ^{-2/3} k^{-7/3}. 
\end{equation}
This result agrees with the numerical result in Fig. 11, where the blue line denotes $k^{-7/3}$. 
The same assumption has been used for classical turbulence; the wave number region that obeys the $-5/3$ power law is known as the inertial range since it originates from the 
inertial term in the Navier--Stokes equation \cite{Frisch}. 
However, in our system, the inertial term is not important for energy transfer, so that the region obeying the power law in Fig. 11 cannot be termed the inertial range. 

The power law in the spectrum of the spin-dependent interaction energy originates from the fact that the energy flux induced by the first term of the right-hand side of Eq. (28) is independent of the wave number.
This nonlinear term contains the second derivative and it differs from the first-derivative inertial term in the Navier--Stokes equation. 
The different nonlinear terms in the two equations are responsible for the different exponents of the energy spectrum. 
Therefore, the $-7/3$ power law of the spectrum is peculiar to turbulence of a spin-1 spinor BEC.

Figure 13 shows how the $-7/3$ power spectrum develops. 
In the early stages of the instability, the spectrum has a peak corresponding to the most unstable wave number of the dispersion relation given by Eqs. (18) and (19), as shown in Fig. 13(a). 
This is confirmed by Fig. 2, from which it follows that the most unstable wave numbers $k \xi$ is approximately equal to $0.3 \sim 0.4$. 
Actually, the density modulation at $t/\tau = 125$ has a stripe structure that resembles those in Figs. 5(a) and 9(a), and the unstable wave number corresponds to the wave number of the stripe. 
After the instability occurs, the spectrum changes to that in Fig. 13(b). 
The density then no longer sustains the stripe structure and excites modulation of various wave numbers. 
As time increases, the spectrum starts to obey the $-7/3$ power law, as shown in Fig. 13(c). After that, the spectrum continues to obey 
this power law. This is confirmed by Fig. 13(d) showing the spectrum at $t/\tau = 5000$.

We calculate the time dependence of the power exponent by the least-squares method \cite{least, size}. 
The deviation $\sigma$ from the straight line obtained by the method is also calculated. 
The results with $V_{R}/c_{s} = 0.39$, $0.78$, and $1.96$ are shown in Fig. 14, which exhibits that the power exponent is approximately $-7/3$ over a long time.
Note that since our system is not stationary, the power exponent asymptotically approaches $-7/3$ in Fig. 14.
Also, Fig. 14(a) shows how the time development of the power exponent depends on the relative velocity $V_{R}$.
When the relative velocity is small, it takes long time for the value of $n$ to approach $-7/3$. 
This is confirmed by comparing the cases $V_{R}/c_{s} = 0.39$ and $0.78$. 
On the other hand, the spectrum behaves differently when $V_{R}/c_{s} = 1.96$. 
In this case, the value of $n$ approaches $-7/3$ from above (Fig. 14(a)). 
This can be understood from the imaginary part of the dispersion relation in Fig. 2, which shows that 
the unstable region of $\omega _{1,-1}$ is broad for a large relative velocity. 
This means that, unlike the case for a small velocity, the instability contains various wave numbers. 
Thus, in this case, the spectrum in the low wave number region $0<k<k_{s}$ becomes flat in the early stages of the instability, leading to the time dependence of $n$. 

In summary, we observed a $-7/3$ power law in spin-1 spinor BECs with the ferromagnetic interaction. This law is independent of 
the relative velocity $V_{R}$ in our numerical calculations based on the GP equation for $0 < V_{R}/c_{s} < 1.96$. However, if the relative 
velocity is much greater than the sound speed $c_{s}$, this law will not hold because the total particle density $n$ may be inhomogeneous. 

\subsection{Antiferromagnetic interaction case}

\begin{figure}[t]
\begin{center}
\includegraphics[keepaspectratio, width=9cm,clip]{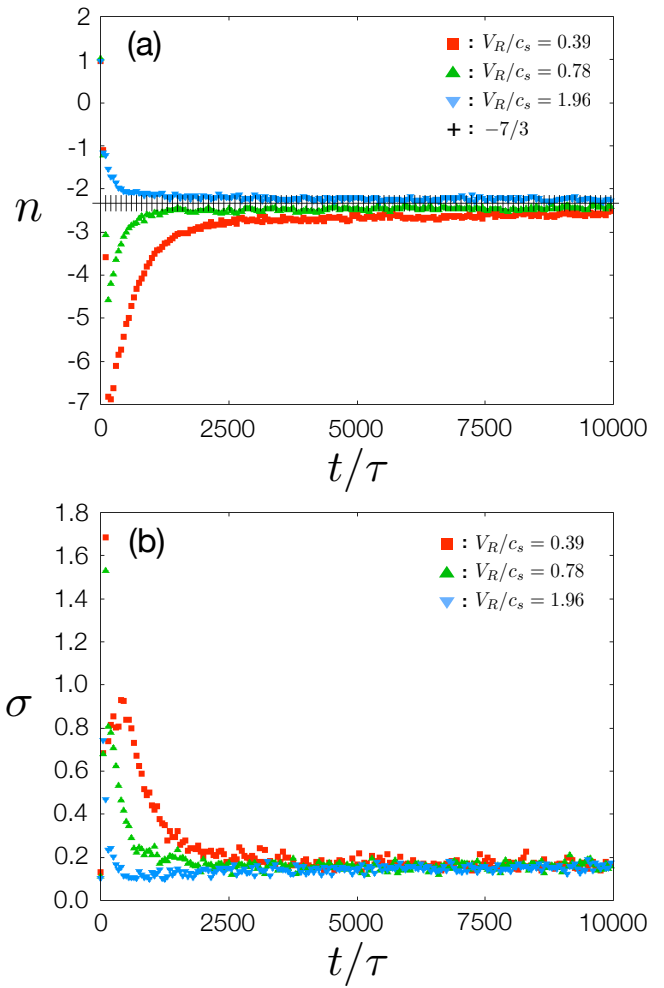}
\caption{Time dependence of the power exponent and deviation from the straight line obtained by the least-squares method for the ferromagnetic case. 
The value of $n$ in (a) is calculated by the least-squares method. 
The deviation $\sigma$ in (b) is defined as $\sqrt{\sum _{i=1}^{N}(y_{i}-f(x_{i}))^{2}/N}$, where $f$, $N$, and $(y_{i}, x_{i})$ are the line obtained by the least-squares method, the number, and the data set. In this calculation, the wave number region ($ k_{0} < k < k_{s}$) is limited whit $k_{0} = 8 \pi /L$. 
This was obtained by performing numerical calculations with $c_{0}/c_{1} = -20$, $c_{0} > 0$, $V_{R}/c_{s} = 0.39, 0.78, 1.96$.} 
\end{center}
\end{figure}

Unlike the ferromagnetic case shown in Fig. 11, the spectrum of the spin-dependent interaction energy in the antiferromagnetic case does not obey the $-7/3$ power law. 
This is confirmed by Fig. 15 which shows the spectrum of the spin-dependent interaction energy for the antiferromagnetic case at $t/\tau = 6000$ when the spin density vector is greatly disturbed.  
However, this spectrum may show a power law in the narrow range $0.3< k\xi <1.4$. 
As shown in the following, we cannot estimate the power exponent by the simple scaling analysis applied to the ferromagnetic case. 

The $-7/3$ power law for the ferromagnetic case was obtained by applying a scaling argument under certain assumptions, which are not applicable for the antiferromagnetic case.
This can be confirmed by checking the PDF of the magnitude of the spin density vector (Fig. 16).
There are two distinct differences between Figs. 12 and 16:
the peak width and the magnitude of the spin density vector corresponding to the peak both differ. 
The peak width in Fig. 16 for the antiferromagnetic case is larger than that in Fig. 12 for the ferromagnetic case. 
The magnitude of the spin density vector corresponding to the peak for the ferromagnetic case is approximately $n_{0}$, whereas that for the antiferromagnetic case is smaller than $n_{0}$. 
These results imply that Eq. (26) is not valid  for the antiferromagnetic case. 
Thus, Eq. (34), which is based on Eq. (26), cannot be applied to the antiferromagnetic case. 
Therefore, the spectrum for the antiferromagnetic case cannot be analyzed by the simple scaling analysis applied to the ferromagnetic case. 

\begin{figure}[t]
\begin{center}
\includegraphics[keepaspectratio, width=9cm,clip]{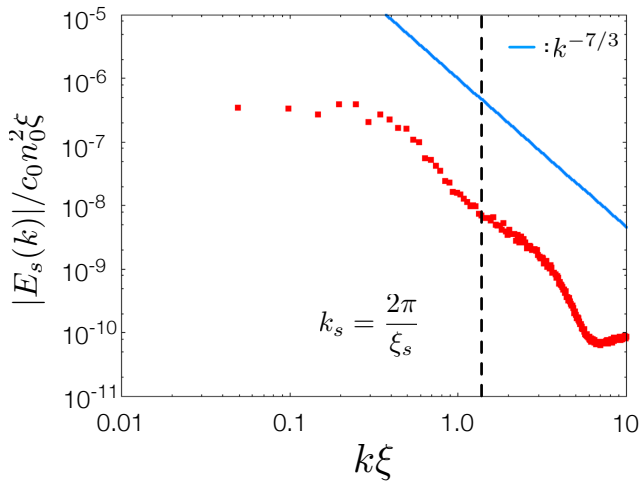}
\caption{Spectrum of spin dependent interaction energy $E_{s}(k)$ for the antiferromagnetic case at $t/\tau = 6000$. 
In the spectrum, the red squares and blue solid and black dashed lines indicate the numerical results for the spectrum, the line proportional to $k^{-7/3}$, and 
the boundary of the wave number corresponding to the spin coherent length, respectively. 
This was obtained by performing numerical calculations with the following parameters: $c_{0}/c_{1} = 20$, $c_{0} > 0$, $V_{R}/c_{s} = 0.78$.} 
\end{center}
\end{figure}

\begin{figure}[t]
\begin{center}
\includegraphics[keepaspectratio, width=9cm,clip]{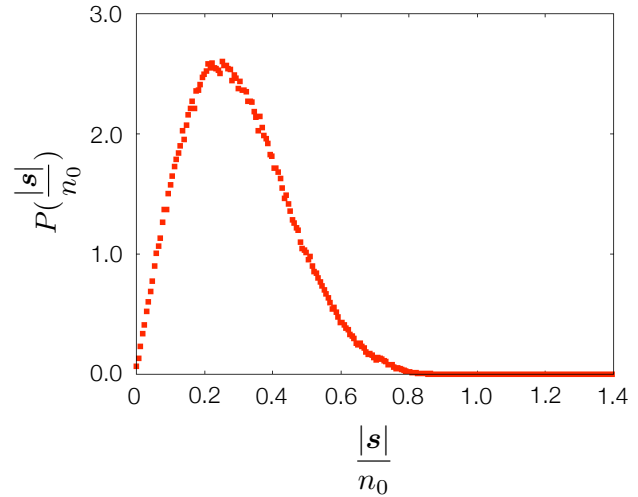}
\caption{Probability density function of the magnitude of the spin density vector for the antiferromagnetic case at $t/\tau = 6000$.
This was obtained by performing numerical calculations with the following parameters: $c_{0}/c_{1} = 20$, $c_{0} > 0$, $V_{R}/c_{s} = 0.78$.} 
\end{center}
\end{figure}

\section{Discussion}
In this section, we discuss the following three topics: (i) comparison between turbulence in spin-1 spinor BECs and other kinds of turbulence; (ii) possibility of observing the $-7/3$ power law, and 
(iii) how to generate counterflow in spin-1 spinor BECs. 

\subsection{Comparison of turbulence in spin-1 spinor BECs and other kinds of turbulence}
We compare the turbulence in spin-1 spinor BECs with that in one-component BECs and classical incompressible fluids.
Through this comparison, we discuss some characteristic properties of turbulence in spin-1 spinor BECs. 

There are some differences between turbulence in one-component BECs and that in spin-1 spinor BECs. 
In the former system, the Kolmogorov $-5/3$ power law spectrum is obtained only if an external force is applied. This force injects energy on large scales and generates quantized vortices. 
If the external force is weak and the number of vortices is not enough, the incompressible kinetic energy 
decreases and the Kolmogorov spectrum tends to disappear \cite{Kobayashi05, Kobayashi05jpsj}. 
Thus, the existence of the quantized vortices is considered to be important for sustaining the Kolmogorov $-5/3$ power law in the turbulence of one-component BECs. 
On the other hand, the $-7/3$ power law in spin-1 spinor BECs is sustained without the application of an external force, as confirmed by Fig. 14. 
In the ferromagnetic case (see Figs. 8 (a) and 10 (a)), the absolute value of the spin-dependent interaction energy of Eq. (22) tends to increase without the application of an 
external force due to the ferromagnetic interaction. 
This is in contrast with the reduction in the incompressible kinetic energy in the turbulence of one-component BECs when no external force is applied. 
We conjecture that this increase allows the system to obey the $-7/3$ power law even when no external force is applied. 

The self-similarity in wave number space can be strongly related to the structures in real space, which is very 
important for understanding the behavior of turbulence in real space. In the turbulence of the one-component BEC, the 
Kolmogorov $-5/3$ power law is believed to be related to the Richardson cascade of quantized vortices, where larger 
vortices become smaller through reconnections of vortices. This cascade may correspond to the energy cascade in 
wave number space, where the incompressible kinetic energy is transported to a high wave number. 
We expect that there are some spin structures characteristic of turbulence with the $-7/3$ power law in a spin-1 spinor BEC. 
Like the Richardson cascade of quantized vortices, some larger spin structures may become smaller to transport the energy to a high wave number. 
However, we currently do not know what kinds of spin structures are essential for the $-7/3$ power law.

We compare turbulence in spin-1 spinor BECs with that in classical fluids. 
In classical turbulence where the Reynolds number becomes infinite, the inertial term of the Navier--Stokes equation 
becomes dominant in the inertial range, which causes the system to obey the $-5/3$ power law. 
We do not know any quantities that correspond to the Reynolds number in spin-1 spinor BECs. However, as our system becomes 
turbulent, the term on the right-hand side of Eq. (30) becomes dominant. 
Hence, the spin-dependent interaction energy in the turbulence is transported to a high wave number mainly by this term; this differs from the turbulence mechanism in incompressible classical fluids. 
This is very significant for the $-7/3$ power law, as pointed out in Sec. V. 

From the above discussion, it follows that the turbulence in spin-1 spinor BECs has some properties that other kinds of turbulence do not have. 
There are several unknowns associated with this turbulence such as the kinetic energy spectrum, the spin structure in real space, the interaction between the velocity and the spin field. 
These will be studied in the future.

\subsection{Possibility of observing the $-7/3$ power law}
The possibility of observing the $-7/3$ power law is discussed. 
We consider that the $-7/3$ power law may be experimentally observed if the following three conditions are satisfied. 

The spectrum of trapped systems with the ferromagnetic interaction is expected to manifest the $-7/3$ power law less clearly than that the spectrum of homogeneous systems.
This is because we neglect the term containing $\bm{\nabla} n$ in Eq. (28) in our derivation of the $-7/3$ power law.
Variation of the total particle density $n$ should affect the spectrum in trapped systems. 
However, for large trapped systems, this is expected to have a small effect on the spectrum. 
Therefore, large trapped systems are preferable for observing the $-7/3$ power law. 
This is the first condition. 

All components of the spin density vector have been experimentally observed by a phase contrast imaging method \cite{Sadler06}.
The expressions for the spectrum of the spin-dependent interaction energy show that it is possible to obtain a spectrum if the spin density vector is observed everywhere. 
This is because Eq. (25) contains only the Fourier component of the vector $\tilde{\bm{s}}(\bm{k})$. 
Thus, the second condition is the observation of the spin density vector.

The experimental resolution in wave number space is important for observing the $-7/3$ power law. 
Our numerical calculations reveals that the spectrum of the spin-dependent interaction energy obeys the $-7/3$ power law in the low wave number region ($k<k_{s}$), as shown in Fig. 11. 
In trapped systems, we expect that the $-7/3$ power law holds in the region $k_{R}<k<k_{s}$, where $k_{R}$ is the wave number corresponding to the system size. 
Typically, the system size is approximately equal to the Thomas--Fermi radius. Thus, experiments for observing the $-7/3$ power law 
need to have a resolution up to the spin coherence length $\xi _{s}$ in real space. 
Since Sadler $et$ $al$. \cite{Sadler06} observed a polar core vortex, the resolution of their experiments seems to be sufficient for observing the $-7/3$ power law. 
The third condition is the observation with a resolution up to the spin coherence length $\xi _{s}$. 
Note that the larger a system size is, the broader the region that obeys the $-7/3$ power law will be, because $k_{R}$ is inversely proportional to the system size. 
Thus, larger systems are more suitable for observing the $-7/3$ power law (this is also consistent with the first condition). 

In summary, we conclude that the $-7/3$ power law may be experimentally observed if the above three conditions are satisfied. 

\subsection{How to generate counterflow in spin-1 spinor BECs}
We discuss the two methods for generating counterflow of spin-1 spinor BECs in trapped systems. 

The first method is to use a double well potential. A numerical study has already investigated this \cite{Guilleumas08}. 
In this case, the $m=\pm 1$ components are separately trapped in each well. The central barrier between two wells is then removed, which generates counterflow in 
spin-1 spinor BECs. 

The second method is to utilize a magnetic field gradient. This method was used to generate counterflow of two-component BECs \cite{Ishino10}. 
In the initial state, the $m=\pm 1$ components are trapped in a harmonic trap. As a magnetic field gradient is applied, one component moves in the 
the field gradient direction, while the other component moves in the opposite direction. 
The magnetic field gradient is switched off when the two components are sufficiently separated, which causes the $m= \pm 1$ components to move toward the center of the trap, generating a counterflow. 

In future, we may use these methods to study counterflow in spin-1 spinor BECs in a trapped system. 

\section{Conclusions}
This paper addressed two main topics: the dynamics induced by counterflow of the spin-1 spinor BECs in a homogeneous two-dimensional system and 
the turbulence generated by the counterflow. These themes are investigated using the GP and BdG equations. 

The results reveal that the properties of the dynamics and turbulence in this system are strongly dependent on whether the spin-dependent interaction is ferromagnetic or antiferromagnetic. 
We summarize the results below.

The dynamics induced by the counterflow in the spin-1 spinor BEC was investigated by performing analytical calculations using the BdG equation and numerical calculations using the GP equation. 
We obtain the dispersion relations of Eqs. (18) and (19) from the BdG equation; these relations show the dynamical instability. 
The dispersion relations depend on the spin-dependent interaction, so that $\rm{Im} [\omega _{1,-1}]$ and $\rm{Im} [\omega _{0}]$ for the 
antiferromagnetic case differ from those for the ferromagnetic case (Figs. 1 and 2). 
The numerical calculations reveal that, in the early stages of the instability, the dynamics can be understood in terms of the dispersion relations. The stripe width of the 
density modulation in Fig. 5(a) and 9(a) is approximately equal to the most unstable wavelength  obtained by the dispersion relations. 
In addition, the isotropy and anisotropy of the density modulation in Figs. 3(b) and 5(a) can be explained in terms of the dispersion relations.
The distinct difference between the ferromagnetic and antiferromagnetic cases appears in the magnitude of the spin density vector, as shown in Figs. 3 $\sim$ 10. 
This is very important for the spectrum of the spin-dependent interaction energy in the turbulence, as pointed out in Sec. V.
These results reveal dynamics peculiar to the spin degrees of freedom. 

We studied the turbulence generated by the counterflow in spin-1 spinor BECs by performing numerical calculations using the GP equation and scaling analysis.
In the ferromagnetic case, the spectrum of the spin-dependent interaction energy obeys the $-7/3$ power law (Fig. 11). The power exponent $-7/3$ is obtained by 
the scaling analysis of Eq. (31) when we make the following three assumptions: the wave functions $\psi _{m}$ are approximately expressed by Eq. (26), 
Eq. (31) are invariant under the scale transformation, and the flux of the spin-dependent interaction energy is independent of the wave number. 

On the other hand, for the antiferromagnetic case, the spectrum does not exhibit the $-7/3$ power law. 
This is probably attributable to the breakdown of the validity of Eq. (26). In this case, the PDF of the magnitude of the spin density vector (Fig. 16) 
differs from that for the ferromagnetic case (Fig. 12), so that such scaling analysis used for the ferromagnetic case is not applicable. 
However, in the narrow range of the wave number, the spectrum in Fig. 15 may show a power law, which cannot be estimated by the scaling analysis. 

These results are characteristic of the spin degrees of freedom and are thus not exhibited by turbulence in one-component BECs and classical fluids. 
There are some unresolved problems associated with turbulence in spin-1 spinor BECs has (see Sec. VI); we intend to study these in the near future.

\section*{ACKNOWLEDGMENT}
M. T. acknowledges the support of a Grant-in-Aid for Science Research from JSPS (Grant No. 21340104).


\begin{thebibliography}{99}
\bibitem{least} The value of $n$ in Fig. 14 was calculated by the least-squares method in the wave number region $k_{0}<k<k_{s}$, where $k_{0}$ is $8\pi /L$ for a system size $L=128\xi$. 
We employ this method irrespective of whether the spectrum exhibits a power law or not. Thus, the value of $n$ in the early stages of the instability 
does not necessarily reflect the actual power exponent because the deviation $\sigma$ is large. 
The values of $n$ are meaningful as a power exponent when the spectrum exhibits a power law (e.g., when $t > 2500$ due to the small deviation $\sigma$).
\bibitem{size} Let us mention the finite size effect which can affect the power exponent. We observe that, in the least square method, the deviation of $n$ 
from $-7/3$ becomes large when $k_{0}$ is smaller than $8\pi /L$. This may be caused by the finite size effect. 
Thus, in Fig. 14, we use $k_{0}=8\pi /L$ to reduce this effect.
\end{thebibliography}
\end{document}